\documentclass[a4papfleqn,usenatbib,useAMS,usedcolumns]{mnras}

\usepackage{subcaption}
\usepackage[T1]{fontenc}
\usepackage{ae,aecompl}
\usepackage{comment}
\usepackage{threeparttable}
\usepackage{booktabs, caption, makecell}

\usepackage{graphicx}	
\usepackage{amsmath}	
\usepackage{gensymb}
\usepackage{siunitx}
\usepackage{comment}

\newcommand{\ie}{{\textrm i.e.}}

\newcommand{\civ}{\mbox{C\,{\sc iv}}}

\newcommand{\siiv}{\mbox{Si\,{\sc iv}}}

\def\h2{$\rm H_2$}
\def\Nh2{$N$(H${_2}$)}

\def\kms{km\,s$^{-1}$}

\def\zem{$z_{\rm em}$}

\def\21{21-cm}

\def\t0{T$_{0}$}

\def\c21{$C_{21}$}

\def\civ{C~{\sc iv}}
\def\siiv{Si~{\sc iv}}

\def\J13{J$1322+0524$}

\def\dw{$\Delta W$}
\def\fdw{$\frac{\Delta W}{W}$}
\def\afdw{\big{|}$\frac{\Delta W}{W}$\big{|}}
\def\mbh{$M_{BH}$}
\def\lbol{$L_{bol}$}
\def\redd{$\lambda_{Edd}$}

\title[UFO sample]{Time variability of ultra-fast BAL outflows using SALT: \civ\ absorption depth based analysis \thanks{Based on observations collected at Southern African Large Telescope (SALT; Programme IDs 2015-1-SCI-005, 2018-1-SCI-009, 2019-1-SCI-019 and 2020-1-SCI-011) and the European Organisation for Astronomical Research in the Southern Hemisphere under ESO programme 093.A-0255.}}

\author[Aromal et al.]{
P. Aromal$^{1}$\thanks{E-mail: aromal@iucaa.in (PA)},
R. Srianand$^{1}$,
and P. Petitjean$^{2}$
\\
$^{1}$IUCAA, Postbag 4, Ganeshkind, Pune 411007, India\\
$^{2}$ Institut d'Astrophysique de Paris, Sorbonne Universit\'e and CNRS, 98bis boulevard Arago, 75014 Paris, France\\
}

\date{Accepted XXX. Received YYY; in original form ZZZ}

\pubyear{2015}

\begin{document}
\label{firstpage}
\pagerange{\pageref{firstpage}--\pageref{lastpage}}
\maketitle

\begin{abstract}
We probe the small-scale absorption line variability 
using absorption depth based analysis of a sample of  64 ultra fast outflow (UFO) \civ\ broad absorption line (BAL) quasars monitored using the Southern African Large Telescope.
We confirm the strong monotonic increase in the strength of variability with increasing outflow velocity.
We identify regions inside the BAL trough for each source where the normalized flux difference between two epochs is $>$0.1 for a velocity width $\ge$500 \kms (called ``variable regions"). 
We find the total number of ``variable regions" increases with time interval probed and the number of BALs showing variable regions almost doubles from short ($<$2 yrs) to long ($>$2 yrs) time scales.
We study the distributions of variable region properties such as its velocity width, depth, and location.
These regions typically occupy a few-tenths of the entire width of the BAL. Their widths are found to increase with increasing time scales having typical widths of $\sim$ 2000 \kms\ for $\Delta t >$ 2 yr.
However, their absolute velocity with respect to \zem\ and their relative position within the BAL profile remain random irrespective of the time scale probed.
The equivalent width variations of the BALs are strongly dependent on the size and depth 
of the variable regions but are little dependent on their total number.
Finally, we find that $\sim$17\% of the UFO BALs show uncorrelated variability within the BAL trough.

\end{abstract}




\section{Introduction}

Strong outflows from quasars manifest themselves as broad absorption lines (BALs) in their spectra which generally have widths of several 1000 \kms\ and outflow velocities reaching up to several 10,000 \kms\ with respect to the emission redshift (\zem) of the quasar. 
About 10-20 $\%$ of all optically selected quasars show BALs 
\citep[][]{Weymann1991}
and this fraction can be as high as 40$\%$ if dust and other observational biases are properly accounted for \citep[][]{dai2008,allen2011}.
Crucial information about the physical state of these outflows and their
origin can be drawn from the study of their variations with time
\citep{Lundgren2007,Gibson2008, Gibson2010, filiz2012, capellupo2011, Capellupo2012, Vivek2014, Filiz2013,mcgraw2017,rogerson2018,cicco2018,aromal2022, green2023}. 

In a previous paper
\citep[][]{aromal2023} we have studied the time variability (over 7.3 yrs) 
of C~{\sc iv} ultra-fast outflows (UFOs) detected in a sample of 64 C~{\sc iv} broad absorption line quasars (with 80 distinct BAL
components) monitored using the Southern African Large Telescope.
These outflows are characterized by maximum outflow velocities greater than 15000 \kms. 
We found that the fraction of variable BALs in our sample (95\%) is much higher than that found for the general BAL population.
The \civ\ equivalent width variations are found to depend on the BAL properties such as its equivalent width  (W), velocity width and absorption depth, etc., but the dependence on quasar properties such as \mbh, \lbol\ and \redd\ are found to be weak \citep[see also][]{Filiz2013}.  
BALs with low W, high velocity, shallow profile, and small velocity width tend to show more variability. When multiple BAL components are present, a correlated variability is often seen between low- and high-velocity components with the latter showing larger variation
amplitude. 
We also find an anti-correlation between the 
variations in the continuum flux and W \citep[see also][]{vivek2019}. While this suggests photoionization-induced variability, the scatter in the continuum flux is found to be much smaller than that of W.

However, while analyzing variability using the total equivalent width
calculated over the entire BAL, which is usually the method adopted in much 
of the literature, we tend to miss out the small-scale variations 
that span a few tenths of the width of the entire BAL
as noted by a few variability studies \citep{Gibson2008, capellupo2011,Filiz2013}.
These local variations may be related to a multi-streaming flow or inhomogeneities in an otherwise smooth flow. 
For example, hydrodynamical disk wind simulations by \citet{proga2000} demonstrated the formation of dense knots from Kelvin-Helmholtz instabilities which propagate along the line-driven fast wind streams.
These knots were seen to be produced every $\sim$ 3 yr in their simulations and the density contrast between the knots and the rest of the wind was suspected to show spectral signatures although this requires more detailed studies. Pixel optical depth based variability studies can be very useful in identifying such inhomogeneous regions and their time evolution. Such measurements provide important constraints on the models.

\civ\ absorption profiles derived from hydrodynamical disk wind simulations show large variability, especially at high velocities
\citep{proga2012}. 
This excess variability at large velocities is attributed to the emergence of fast mass ejections at relatively large distances where they are well shielded from X-ray radiation.
Note that the variability seen in these simulated \civ\ BAL-like profiles almost always leads to highly saturated absorption and cannot reproduce the diverse nature of profile variability seen in observations.
However, these simulations indicate that even though absorption happens over a large velocity range, the optical depth variations may be non-uniform throughout the profile.
Hence, in this paper, we wish to study the velocity dependence of BAL variations and thus the absorption profile's small-scale variability.
Since our sample focuses on high-velocity outflows, we hope this can help in constraining various physical parameters that are crucial for disk wind simulations.
However one needs to keep in mind that our study is restricted to UFO BALs and it may require further study to extend our results to the general BAL population.

The paper is organized as follows.
In Section~\ref{sec:sample}, we present our UFO sample.
Section~\ref{sec:trans_flux_analysis} provides details of  the absorption depth based analysis and the results on the velocity dependence of BAL variability.
Section~\ref{sec:varregion} presents the method for identifying variable regions inside the BAL and the results on various properties of variable regions and their time dependence.
In Section~\ref{sec:conclusions}, we discuss the main results and provide a summary of our work.
Throughout this paper we use the flat $\Lambda$CDM cosmology with  $H_0$ = 70 \kms\ Mpc$^{-1}$ and $\Omega_{m,0}$ = 0.3. 

\section{BAL-UFO sample}
\label{sec:sample}
Our sample of \civ\ UFO BAL quasars is constructed from the Sloan Digital Sky Survey data release 12 (SDSS DR12) quasar population \citep{paris2017} after applying the following selection criteria.
First, we demand the BAL parameter to be 1 and the Balnicity index (BI) to be greater than 0 \kms\ in the catalog of \citet{paris2017}. 
Then, as per the definition of UFO BALs, the observed maximum outflow velocity with respect to \zem\
at the time of observations is required to be greater than 15000 \kms. 
Note that the emission redshift (\zem) 
is taken from \citet{Hewett2010} (which derives the systemic redshift from the fit to the C~{\sc iii}] emission line) whenever 
available and otherwise from \citet{paris2017}. 
Also, we restrict our sample to quasars having emission redshift $z_{\rm em} > 2.0$ to ensure that the \civ\ (also \siiv\ in most cases) absorption falls in the sensitive wavelength range of both the SDSS telescope and the Southern African
Large Telescope (SALT).
We then restrict our sample to objects with declination $<+10$ deg to ensure that the source is accessible to  SALT and with magnitude brighter than $m_r$= 18.5 mag to obtain a sufficiently high spectroscopic signal-to-noise ratio (SNR) with SALT. 
%
After all these constraints, we end up with a total of 63 sources with UFO BALs from the SDSS DR12 catalog to which we add another UFO BAL source namely J132216.25+052446.3, an interesting BAL quasar we have been monitoring for the past 7 years using SALT \citep{aromal2022}.
Our final sample consists of 64 UFO BAL quasars.
We identify 80 BALs in these sources according to the definition given in \citet{Weymann1991}.
The list of sources, log of observations, and details of spectra obtained at different epochs are given in Table B1 in the online material of \citet{aromal2023}.


\section{absorption depth based analysis}
\label{sec:trans_flux_analysis}

Earlier BAL variability studies \citep{Gibson2008, capellupo2011,Filiz2013} have shown that the BAL trough variability is dominated by  ``variable" regions within the trough.  As discussed above, these variable regions could be manifestations of small scale density and/or velocity inhomogeneities introduced by instabilities in the flow.
In the case of simple 1D disk-wind models \citep[like][]{murray1995} there is a one-to-one correspondence between the density, velocity, and location of the absorbing gas. In such cases, the absorption seen at a given velocity is produced by gas with a definite location.
However, in realistic simulations the velocity field can be complex and flow can have multiple components \citep[see for example,][]{Proga2004}.
Therefore, the absorption observed at a given velocity can be produced
by a mixture of gas at different locations.
Also, in these simulations, it is found that the covering fraction of the radiatively driven quasar outflows is highly dependent on the outflow velocity.
%
%
%
Hence, studying the variability of the absorption over smaller velocity intervals using 
an absorption depth based analysis, which looks at the strength of the absorption quantified by its depth as a function of outflow velocity, can provide us with important clues on the nature and evolution of BAL outflows.

\begin{figure}
    \centering
    \includegraphics[viewport=80 0 2380 720, width=\textwidth,clip=true]{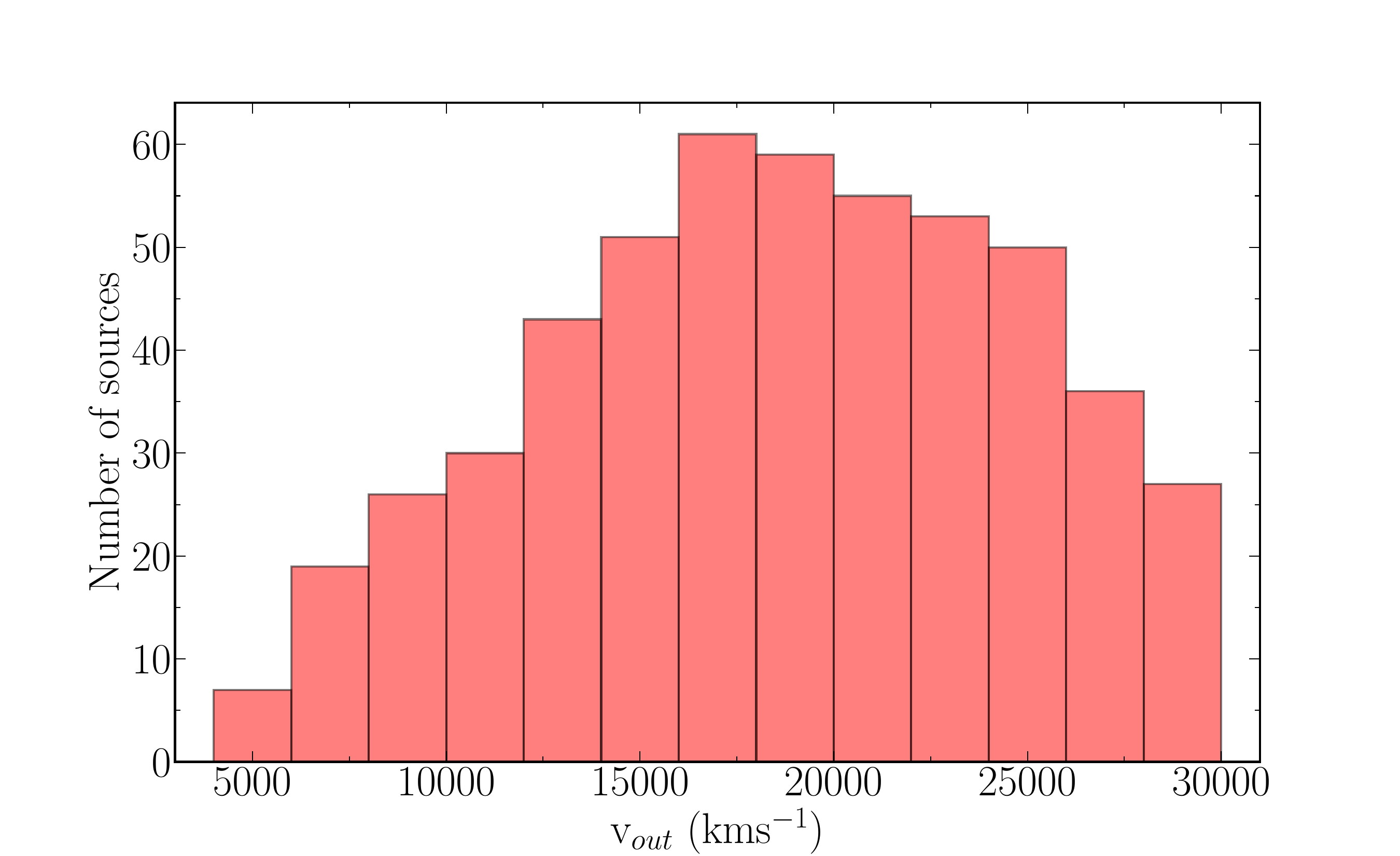}
    \caption{The number of sources contributing to each velocity bin of width 2000 \kms\ is shown starting from 4000 to 30000 \kms\ 
    from the emission redshift taking into account all the UFO sources in our sample. 
    }
    \label{fig:nBAL_vs_velocity}
\end{figure}

In the following, we will consider windows of width 2000 \kms\ which corresponds
to the typical width of coherent absorption usually seen within BALs. We will 
restrict ourselves to the velocity range 4000-30000 \kms\ relative to $z_{em}$
to avoid the associated absorption in the low velocity end and to limit the study to the
wavelength range between the C~{\sc iv} and Si~{\sc iv} emission lines. 
The number of BAL troughs contributing to each velocity bin of width 2000 \kms\ (considering all the UFO sources in the sample)  over the velocity range 4000-30000 \kms\ is shown in Fig.~\ref{fig:nBAL_vs_velocity}. 
We require that the absorption cover the full range of a velocity bin that is being considered. 
As expected from our sample selection, for the velocity range from 12000 \kms\ to 28000 \kms\, at least half of the BAL QSOs in our sample contribute a point. 
Therefore, this will be the main velocity range of interest for our analysis. 

\subsection{Fractional variation in EW of \civ\ 
BAL 
}
\label{sec:fdw_vs_velocity}

\begin{figure}
    \centering
    \includegraphics[viewport=35 10 2450 1200, width=\textwidth,clip=true]{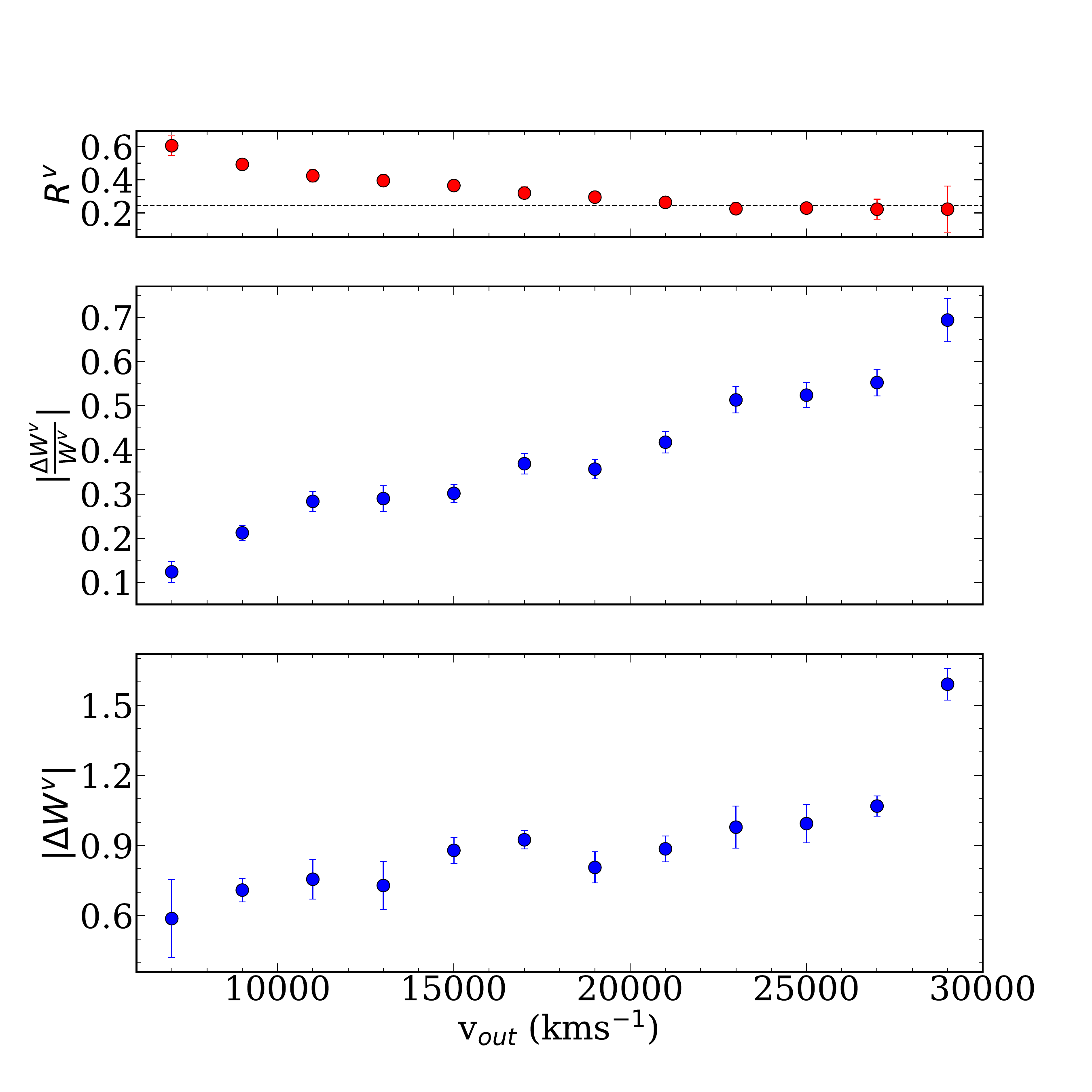}
    \caption{The mean absorbed flux, $R^v$ (top panel), mean fractional change in equivalent width, $\frac{\Delta W^v} {W^v}$, (middle panel) and mean absolute change in equivalent width, $\Delta W^v$  (bottom panel) are plotted as a function of $v_{out}$ in velocity bins of 2000 \kms\ width using the method as mentioned in Section~\ref{sec:fdw_vs_velocity}.
    The horizontal dotted lines in the top panel at $R^v$ = 0.24 show that the points greater than  v$_{out}$ = 18000 \kms\ have almost constant $R^v$ values within error bars even though the corresponding $\frac{\Delta W^v} {W^v}$ values are showing velocity-dependent variations. }
    \label{fig:delw_velocity}
\end{figure}

In \citet{aromal2023} we have seen the BAL components at higher ejection velocities show larger variability in \civ\ equivalent width compared to the components at low ejection velocities. Here we test whether this is true even when we perform velocity bin based analysis without identifying individual BAL components. To do so,
for each BAL profile, we calculate the \civ\ equivalent width in each velocity bin (of velocity width 2000 \kms), denoted as  $W^v$,
and quantified the variability of $W^v$ between two epochs (1 and 2) using,

\begin{equation}
    \Delta W^v = W^v_{2} - W^v_{1}, \quad \sigma_{\Delta W^v} = \sqrt{\sigma_{W^v_1}^2 + \sigma_{W^v_2}^2}
\end{equation}

\begin{equation}
    \frac{\Delta W^v} {W^v} = \frac{W^v_{2} - W^v_{1}}{ (W^v_{1} + W^v_{2}) \times 0.5 }~, 
\end{equation}

\begin{equation}
    \sigma_{\frac{\Delta W^v}{W^v}} = \frac{4 \times (W^v_1 \sigma_{W^v_2} + W^v_2 \sigma_{W^v_1})}{(W^v_1 + W^v_2)^2} \nonumber
\end{equation}
where $W^v_1$ and $W^v_2$ are equivalent widths measured for a particular velocity bin at epochs $t_1$ and $t_2$ respectively with $t_1$ < $t_2$. 
Thus an increase (or decrease) of $W^v$ with time results in a positive (or negative) $\Delta W^v$.
Note that the equivalent width calculated for the entire BAL profile is denoted as W without any superscript. Other notations related to variations in W are same as above without the superscript as done in \citet{aromal2023}.

We estimate $\Delta W^v$ and $|\frac{\Delta W^v} {W^v}|$ for each velocity bin between all possible combinations of epochs with the time separation between them ($\Delta t$) falling in a 2-3.5 yr time bin for all the UFO sources. 
We chose this time bin as it is the most well-sampled time bin in our sample \citep[see][for details]{aromal2023}.
To avoid any bias in the $\Delta W^v$ and  $|\frac{\Delta W^v} {W^v}|$ distributions due to the non-uniform number of available spectroscopic epochs for our sources,
we randomly choose one measurement of
$\Delta W^v$ and  $|\frac{\Delta W^v} {W^v}|$
per source in each velocity bin
and measure the mean for the corresponding
$\Delta W^v$ and  $|\frac{\Delta W^v} {W^v}|$
distribution.
We repeat this procedure 100 times and calculate the mean and $\sigma$ of the resulting distribution.
In Fig.~\ref{fig:delw_velocity}, we plot this mean and standard deviation of the $|\frac{\Delta W^v} {W^v}|$ (middle panel) and $\Delta W^v$ (bottom panel) distributions as a function of the outflow velocity ($v_{out}$). 
This clearly shows the strength of variability as quantified by $|\frac{\Delta W^v} {W^v}|$ increases significantly with increasing $v_{out}$. A similar trend is seen in the case of $\Delta W^v$  as well albeit restricted somewhat to the 
two extreme (smallest and largest) outflow velocities. 

Next, we check if this effect is only due to the comparatively lower optical depth of the absorption as the $v_{out}$ increases.  
For this, we also plot the mean absorbed flux or mean depth of the absorption, $R^v$, ($R^v$ = 1 - $F^v$, where $F^v$ is the mean residual flux) as a function of velocity using the same method in the top panel of Fig.~\ref{fig:delw_velocity}.
Even though we see $R^v$ decreasing with velocity, it is indicated from the figure that this alone cannot account for the increasing $|\frac{\Delta W^v} {W^v}|$, especially at velocities greater than 18000 \kms\ after which the mean $R^v$ remains roughly constant within error bars (as indicated by the black dashed lines in the top panel of Fig.~\ref{fig:delw_velocity}). 
This argument is further supported by the $\Delta W^v$  vs $v_{out}$ plot (bottom panel of Fig.~\ref{fig:delw_velocity}) where $\Delta W^v$  shows an increasing trend with increasing $v_{out}$.
Hence, we can conclude that the increasing fractional variability with velocity is an intrinsic property of these UFO BAL sources and not just a secondary effect due to decreasing equivalent width.

It has been reported by many authors that the high velocity BAL components show large equivalent width variations compared to the low velocity ones \citep{Filiz2013, Vivek2014}.
%
The absorption depth based analysis (by splitting the BAL profile into different velocity bins) allows us to probe whether such a velocity dependence is seen within a BAL profile itself.
Using this approach, our results indicate a strong dependence of variability on the velocity.
In addition, we checked if these results are sensitive to (a) the width of the velocity bin assumed and (b) the chosen variability time scale. To address the former point, we used 1000 \kms\ velocity bins and for the latter, we considered other time bins such as $\Delta t < 2$ yr to redo the analysis mentioned above. We obtained consistent results where variability, i.e. $|\frac{\Delta W^v} {W^v}|$ is a strongly increasing function of velocity irrespective of these choices.

This approach has also been carried out by \citet{Gibson2008} on a relatively smaller sample and found no correlation between BAL variability and the outflow velocity.
Also, \citet{capellupo2011} carried out a similar study on a sample of 24 luminous high redshift quasars and found that the mean depth variations of \civ\ BAL after splitting them into bins of 1000 \kms\ width show no dependence on the velocity. However, their results suffer from the small sample size and less number of epoch pairs used for the analysis. The difference could also come from the fact that the mentioned sample has a diverse population of BALs whereas we mainly focus on UFO BALs.

\begin{figure}
    \centering
    \includegraphics[viewport=50 20 2420 900, width=\textwidth,clip=true]{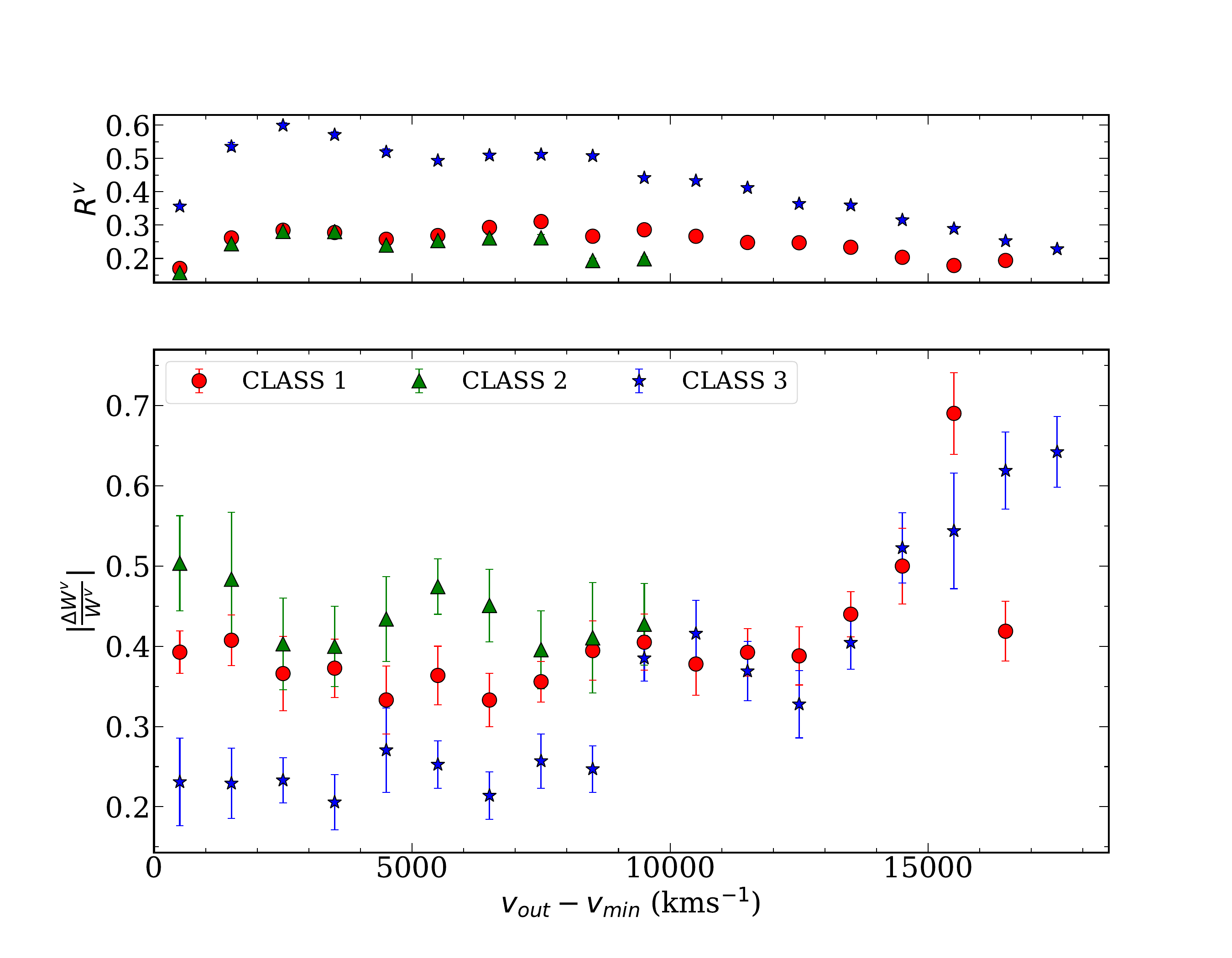}
    \caption{The mean absorbed flux, $R^v$ (top panel), mean fractional change in equivalent width, $|\frac{\Delta W^v} {W^v}|$ (bottom panel) are plotted as a function of relative outflow velocity with respect to the minimum velocity of the BAL ($v_{out}- v_{min}$) in velocity bins of 1000 \kms\ width using the method as mentioned in Section~\ref{sec:fdw_vs_velocity}.
    }
    \label{fig:delw_velocity_classes}
\end{figure}

In \citet{aromal2023}, we probed the velocity-dependent equivalent width variability in three subsamples (Class 1, 2, and 3) defined based on the global absorption profile structure (see figure 1 therein).
In short,
Class 1 includes sources with one or more \civ\ UFO BALs having $v_{\rm min}>8000$ \kms, Class 2 consists of only sources with multiple BAL troughs having at least one UFO BAL with $v_{\rm min}>8000$ \kms\ and one non-UFO BAL with $v_{\rm min}<8000$ \kms\ and finally Class 3 contains sources with a single UFO BAL trough having $v_{\rm min}<8000$ \kms.
Now, we explore how the profile changes with respect to the minimum velocity of the BAL depending on the nature of the profile shape as defined above.
In the bottom panel of Fig.~\ref{fig:delw_velocity_classes}, we present the results of $|\frac{\Delta W^v} {W^v}|$  for the three different classes. In the top panel of this figure, we show the mean absorbed flux, $R^v$, 
as a function of velocity with respect to the minimum velocity of the BAL. It is evident
from the figure that for Class 1 and Class 3 where the \civ\ absorption spans a wider range in velocity, we do see a clear trend of increasing $|\frac{\Delta W^v} {W^v}|$ with velocity.  At the lower velocity ranges Class 3 objects show lower variability compared to the absorption from the other two classes of UFO BALs. 
As in \citet{aromal2023} we can attribute this to slightly larger saturation in this case (see the top panel of Fig.~\ref{fig:delw_velocity_classes}). In this velocity range, we also notice the absorption from Class 2 UFO BALs to show slightly larger variability compared to that of Class 1. 
Interestingly the $R_v$ values for both the classes are nearly the same as shown in the top panel of Fig.~\ref{fig:delw_velocity_classes}.
All these findings are consistent with the results derived by \citet{aromal2023} using equivalent width analysis.

\section{Variable regions and their properties}
\label{sec:varregion}
Even though the BALs in our sample spread over a few  thousand to almost a few tens of thousand \kms, it is clear from visual 
inspection of the spectra that the widths of the regions with the most important \civ\ absorption variability inside the BAL 
are generally much smaller than the total width of the BAL. 
We wish to identify these regions inside the BALs where significant variability occurs. As in the literature, we refer to them as ``variable regions". \citet{Filiz2013} defined the quantity $N_{\sigma} = \frac{F_2-F_1}{\sqrt{\sigma_1^2 + \sigma_2^2 }}$ where $F_1$ and $F_2$ are the normalized flux and $\sigma_1$ and $\sigma_2$ are the corresponding uncertainties per pixel in the spectra for two epochs. They  defined a variable 
region as velocity windows over which 
absorption is detected with $N_{\sigma} \geq 1$ or 
$N_{\sigma} \leq -1$ 
(to take care of both the increasing and decreasing transmitted flux with time)
for at least five consecutive data points (i.e over a typical 
width of more than 275 \kms).
But we plan to put more stringent conditions on variable regions compared to \citet{Filiz2013} since this will help us characterize 
regions with large variability inside a BAL better.
This is partly motivated by the stronger BAL variability shown by our UFO sample and also to alleviate the possible uncertainties that arise from continuum placement.
We believe these large variability regions will provide important clues on the dynamics and evolution of the BAL outflow and also shed light on the nature of small scale inhomogeneities  and how they drive the large equivalent width changes one usually measures in different BAL components.

We define ``variable regions" as those regions inside the BAL trough where the flux difference ($\Delta F = F_2-F_1$ ) between two epochs is greater than 0.1 as a conservative lower limit in
every pixel over a velocity width of at least 500 \kms (i.e. typically 10 consecutive pixels). 
This choice is also motivated by the histogram of the strength of variations in variable regions studied by \citet{Gibson2008}. Given the typical SNR achieved in our spectra \citep[i.e $\ge10$ as can be seen from the last column of table B1 of][]{aromal2023} we usually have $N_\sigma \ge3$ per pixel. This ensures that $N_\sigma \ge10$ for the overall regions identified as ``variable regions" in our analysis.
%
%

\begin{figure}
    \centering
    \includegraphics[viewport=50 20 2400 800, width=\textwidth,clip=true]{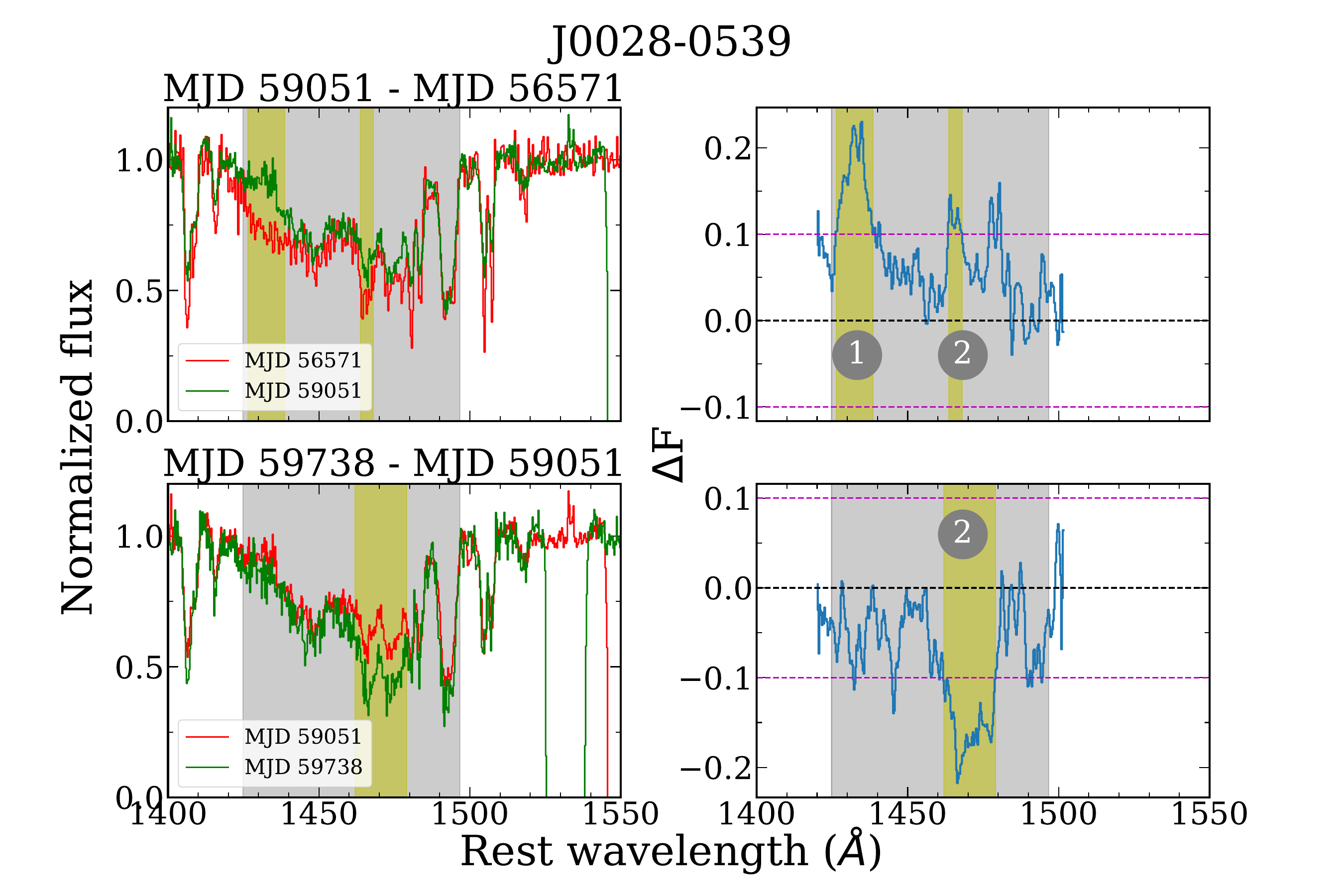}
    \caption{An example of the identification of variable regions in J0028-0539 (shown as yellow shaded regions) inside the BAL (gray shaded region) with spectra and flux difference shown in the left and right panels respectively for two consecutive pairs of epochs. The upper and lower panels correspond to different pairs 
    with MJDs shown above the left panels and legends as well.
    The two identified variable regions are marked '1' and '2' in the first epoch pair. Variable region '2' is observed in the second epoch pair as well with a change in its velocity width. Interestingly the sign of the optical depth variations also change for variable region '2'.
    This indicates the number of variable regions and the width of the variable regions can change when different consecutive epoch pairs are considered.
    }
    \label{fig:eg_var_region}
\end{figure}

We adopt two methods for identifying the location of variable regions 
for each source,
(i) for a given time range, we consider the pair of epochs for which the BALs show maximum \afdw\
and identify variable regions from these two epochs, or
(ii) we consider all consecutive pairs of epochs with time separation  ($\Delta t$) falling into a specific time range as mentioned in the 
analysis and identify the variable regions from these pairs. 
Hereafter we  refer to them as  method I and method II respectively. While method I ensures the analysis is not biased by the non-uniform number of spectroscopic epochs for each quasar, method II allows us to study the evolution of variable regions in individual BAL troughs across the epochs in terms of its position, strength and width.
Note that, in many cases, there can be more than one variable region inside the same BAL trough for a given pair of epochs. 
An example of this procedure is shown in Fig~\ref{fig:eg_var_region} 
in case of one of the sources (J0028-0539) in our sample.



%

\subsection{Properties of variable regions}

After identifying the variable regions using the methods described above, we characterize them using six quantities which are defined as follows:

\begin{enumerate}
    \item $\Delta v_{var}$ : The velocity width of the individual variable region identified between two epochs.  When multiple variable regions are identified in a single BAL trough, all of their widths are considered separately.
    \item f$_{var}$ : The fraction of the total width of the variable regions, i.e., the sum of $\Delta v_{var}$ of all the individual variable regions detected between two epochs divided by the total velocity width of the BAL trough.
    \item  $d_{var,max}$ : The maximum observed depth in the individual variable region considering each pixel in the spectra of the pair of epochs. This quantity will tell us about the level of saturation and possible indications of the average covering factor of the outflow.
    \item $\Delta d_{var,max}$ : The maximum change in the depth of the variable region between the epochs. We calculate the difference in depth pixel by pixel and find the maximum value. 
    \item $v_{var,mid}$ : The mid-velocity of the individual variable region between the epochs considered, \ie\ $v_{var,mid} = \frac{v_{var,min} + v_{var,max}}{2}$, where $v_{var,min}$ and $v_{var,max}$ are the minimum and maximum velocities of the individual variable regions with respect to $z_{em}$.
    \item $l_{var}$ : The relative position of the individual variable region with respect to the BAL defined as  $l_{var}$ = $\frac{v_{var,mid} - v_{min}}{v_{max} - v_{min}}$ where $v_{min}$ and $v_{max}$ are the minimum and maximum velocity of the BAL under consideration relative to $z_{em}$. This means that an individual variable region is more towards the low velocity side of the BAL if $l_{var}$ is close to zero,  and similarly more towards the high velocity side if $l_{var}$ is close to one. 
\end{enumerate}

\subsection{Occurrence statistics of variable regions}
\begin{figure}
    \centering
    \includegraphics[viewport=80 0 2380 600, width=\textwidth,clip=true]{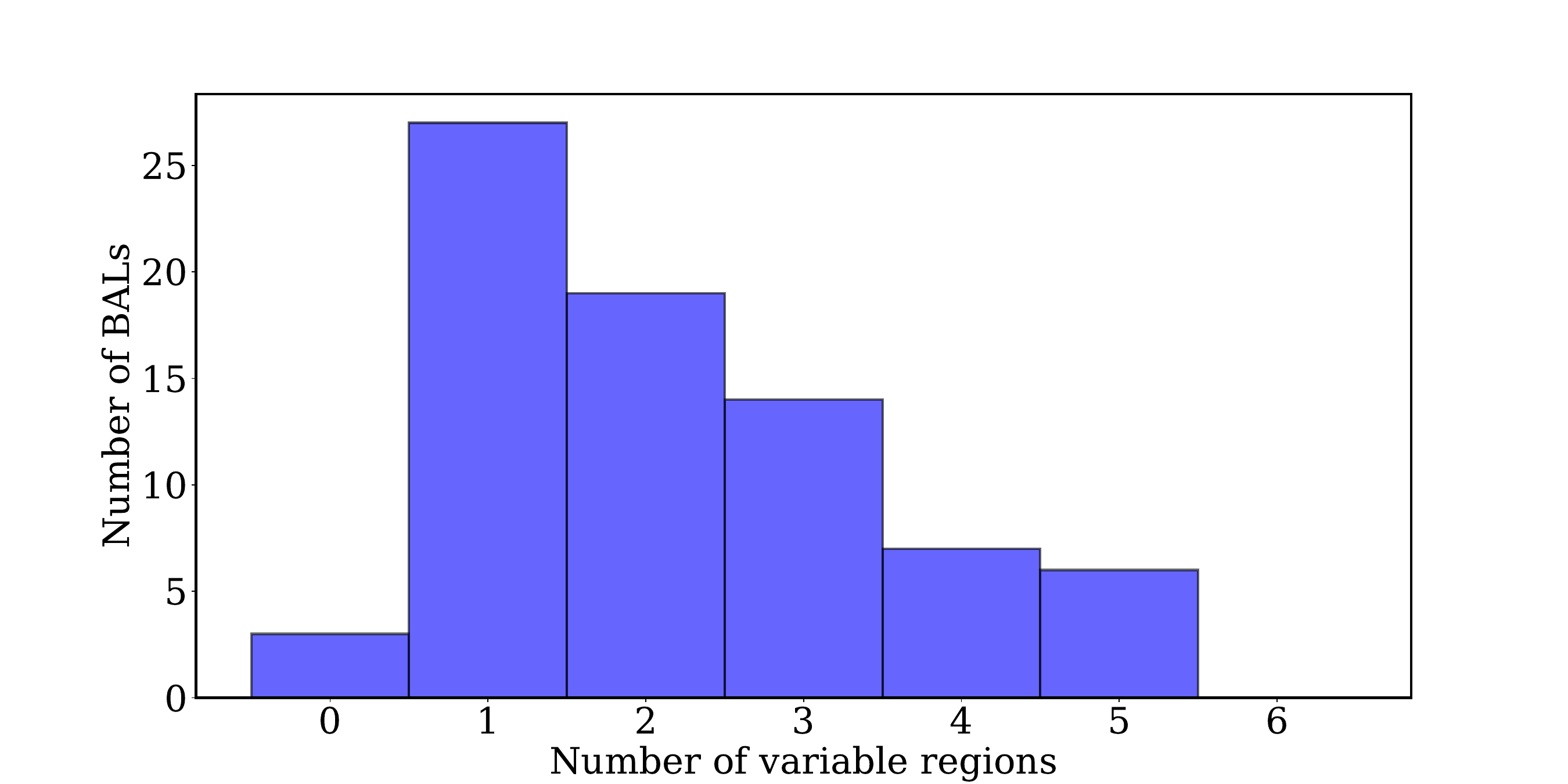}
    \caption{The number of BALs having different number of variable regions using method 1.}
    \label{fig:no_of_var_region}
\end{figure}

In \citet{aromal2023} we have shown that $95$\% of the BAL troughs in our sample do show variable BAL absorption. The distribution of the number of BAL quasars having a given number of variable regions is shown in Fig.~\ref{fig:no_of_var_region}. 
Note that here we consider the one epoch pair per quasar where the maximum change in \fdw\ is observed.
The maximum number of BAL QSOs in our sample tends to have only one variable region.
In the present sample, 61 of the 64 BAL quasars show at least one variable region in their BAL trough. If we confine ourselves to the BAL QSOs showing detectable equivalent width variations, we note that 60 of the 61 variable BALs do show at least one detectable variable region.
This confirms that nearly all the variable BAL troughs in our sample do show at least one variable region with the definition we adopted. 
In comparison, \citet{Gibson2008} found only 1 out of their 13 BAL QSOs to not show even one variable region. In the case of \citet{Filiz2013} $\sim$88\% of the BAL troughs that are known to produce $>3\sigma$ equivalent width variations do show at least one variable region. 
They also identified variable regions in cases where no significant equivalent width variations are seen. In our case, three objects do not show significant equivalent width variability. 
We do detect a variable region in one of these BAL QSOs. The lack of significant detection in the equivalent width variation seen in this case can be attributed to uncorrelated variability which will be discussed in detail below. 

Next, we investigate the frequency of occurrence of variable regions in two time intervals [i.e $<$2 yrs (referred to as short) and $>$2 yrs (referred to as long that typically spread over 2$-$7.3 yrs)] using method I. 
This will allow us to study the occurrence and various properties of variable regions as a function of the time separation between two epochs.
We find 82 and 166 variable regions from 42 and 73 BALs respectively for the short and long time-scales (keep in mind that in some cases, one quasar has multiple BALs in our sample). 
This is on average $\sim$1.95 and $\sim$2.27 variable regions per BAL probed over these time scales. 
These numbers are consistent with the mean $\sim$2.31 found by \citet{Gibson2008} for their 13 BALs studied over a time scale of 3-6 yrs. In the case of \citet{Filiz2013} this ratio is $\sim2.11$ for their sample with time-scale probed greater than 1 yr. In \citet{aromal2023}, we have reported that the UFO BALs in our sample show larger variability compared to the BAL QSO samples discussed in \citet{Filiz2013}. However, when it comes to the occurrence of variable regions the BAL troughs in both samples seem to behave similarly. However, it is the strength of the variability which is higher in the case of UFO BALs.


There are {41} BALs (in {36} quasars) for which our data allows us to probe the presence of variable regions both over the short and long time scales. The number of variable regions identified in the case of short and long time-scales are {80} and {95} respectively. In the case of {16} (respectively {8}) of these BALs we see the number of variable regions to be higher (respectively lower) for the long time scale compared to the small time scale. In the case of {17} BALs the number of variable regions does not change between short and long time scales.

Apart from time scales, another important factor that can affect the nature of variable regions is the sign of BAL variations itself. 
The number of BALs showing weakening and strengthening signatures in terms of their total equivalent width are 48 and 32 respectively out of 80 BALs using method I.
One would expect that this asymmetry may lead to a possible bias in our analysis as the variability is limited to the W of the BAL region in the case of weakening whereas the limit can go as far as saturation in the case of strengthening.
Additionally, the same effect can also lead to differences in depth variations as well.
We explore these aspects in detail in Section~\ref{subsec:var_region_signW}.




\subsection{Common variable regions}
 
\begin{figure}
    \centering
    \includegraphics[viewport=25 40 2400 1720, width=\textwidth,clip=true]{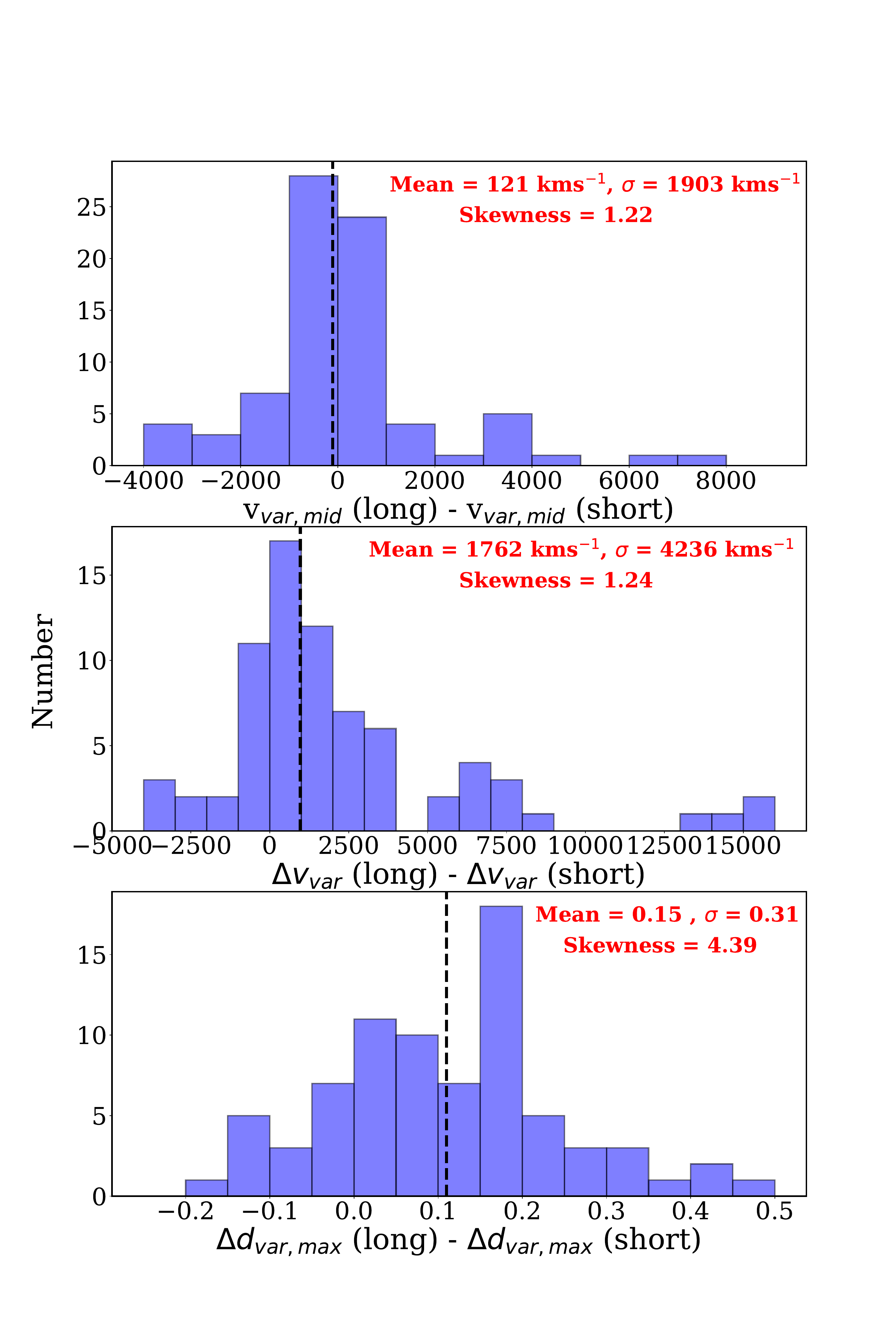}
    \caption{The distribution of the differences in mid-velocity (top),  velocity width (middle) and maximum difference in depth (bottom) of the ``common variable" regions between the short and long time-scales per source.  The vertical dashed lines indicate the median value.
    }
    \label{fig:var_reg_overlap}
\end{figure}

Here we consider only the 41 BALs that have short and long time-scale observations available.
We define ``common variable" regions as regions where there is an overlap in the velocity range between the variable regions identified over the short and long time scales. This definition allows for a possible change in the velocity centroid and the width of the variable regions.

In {38} BALs we find common variable regions that are present both in the short and long time scales when we use method I. In {25} of these cases the width of the common variable region is found to be larger for the long time scale.  While {6} of them show decreasing width with increasing time-scale in {7}  cases there is no significant change in the velocity width.


We notice that in several cases more than one variable region identified over the short time scale  coincides with one variable region identified in the long time scale and vice versa.  
In the top panel of Fig.~\ref{fig:var_reg_overlap}, we plot the histogram of the difference between the mid-velocity of the variable regions ($v_{var,mid}$) identified over the short and long time-scales including multiple coincidences. 
The velocity shift is mostly within $\pm$1000 \kms. The distribution is also by and large symmetric around zero. The large shifts we can see are for the cases when there is a large change in the velocity width of the ``common" variable region that also tends to be asymmetric.

In the middle panel of Fig.~\ref{fig:var_reg_overlap} we show the width difference of the common variable region between the short and long time-scales.
The distribution is skewed (with typical values in the range $-$1000 to $+$2000 \kms) towards positive values indicating an overall increase in the width of the common variable regions with increasing time.  We find the 
common variable regions showing large velocity differences are also the ones showing large velocity width differences.

In the bottom panel of Fig.~\ref{fig:var_reg_overlap} we plot the difference in $\Delta d_{var,max}$ measured at long and short time scales. It is evident from the figure that the distribution is skewed towards the positive values. That is, absorption depth variations in the common variable regions are larger when we consider longer time scales. In summary, in the case of common variable regions, the velocity width and depth of the variable region are found to increase with increasing time scale. This is in line with our finding that the variability amplitude of \civ\ rest equivalent width is higher when probed over longer  time scales compared to short time scales. 

We detect {8} and {31}  variable regions that are seen only in the short or long time scales respectively. This is the complementary set of common variable regions. Such regions are interesting for probing the growth and decay time scale of variable regions. 
A careful investigation of the absorption profile has revealed that these regions do not correspond to distinct velocity components visible in the spectra (i.e. emerging or absorbing velocity components) but often reflect subtle changes in the profile at the locations where we identify variable regions.

{\it The exercise presented here till now for the 41 BALs, that have short and long time-scale observations available, suggests that large optical depth variations occur over smaller velocity ranges. The number of variable regions and their widths increase with increasing time. 
In the case of common variable regions, on average the velocity width seems to increase with time. These regions also show larger absorption depth variations at long time scales.}
In the following section, we explore whether similar trends are seen when we consider the full sample (i.e. including quasars that are not probed over wide range time-scales).



\subsection{Statistical distribution and time dependence}

\begin{table}
    \centering
\caption{Statistics of the variable region parameters}
 \begin{tabular}{ccccccc}
  \hline
  Parameter & \multicolumn{3}{c}{< 2 yr} & \multicolumn{3}{c}{> 2 yr} \\ 
            & 16 & 50 & 84  & 16 & 50 & 84\\
  \hline\hline
  $\Delta v_{var}$ & 618 &1236 &3774 &823&2066 &4097 \\
    f$_{var}$ &0.08 &0.25 &0.60 &0.17 &0.48 &0.83 \\
    $d_{var,max}$ &0.27 &0.44 &0.65 &0.34 &0.59 &0.79 \\
    $\Delta d_{var,max}$ &0.18 &0.25 &0.33 &0.22 &0.32 &0.46 \\
    $v_{var,mid}$ &13426 &19339 &24740 &10807 &17947 &23960 \\
    $l_{var}$ &0.23 &0.59 &0.88 &0.17 &0.51 &0.83 \\

   \hline 
 \end{tabular}

\label{tab_var_prop}
\end{table}

\begin{figure}
    \centering
    \includegraphics[viewport=40 90 2400 1600, width=\textwidth,clip=true]{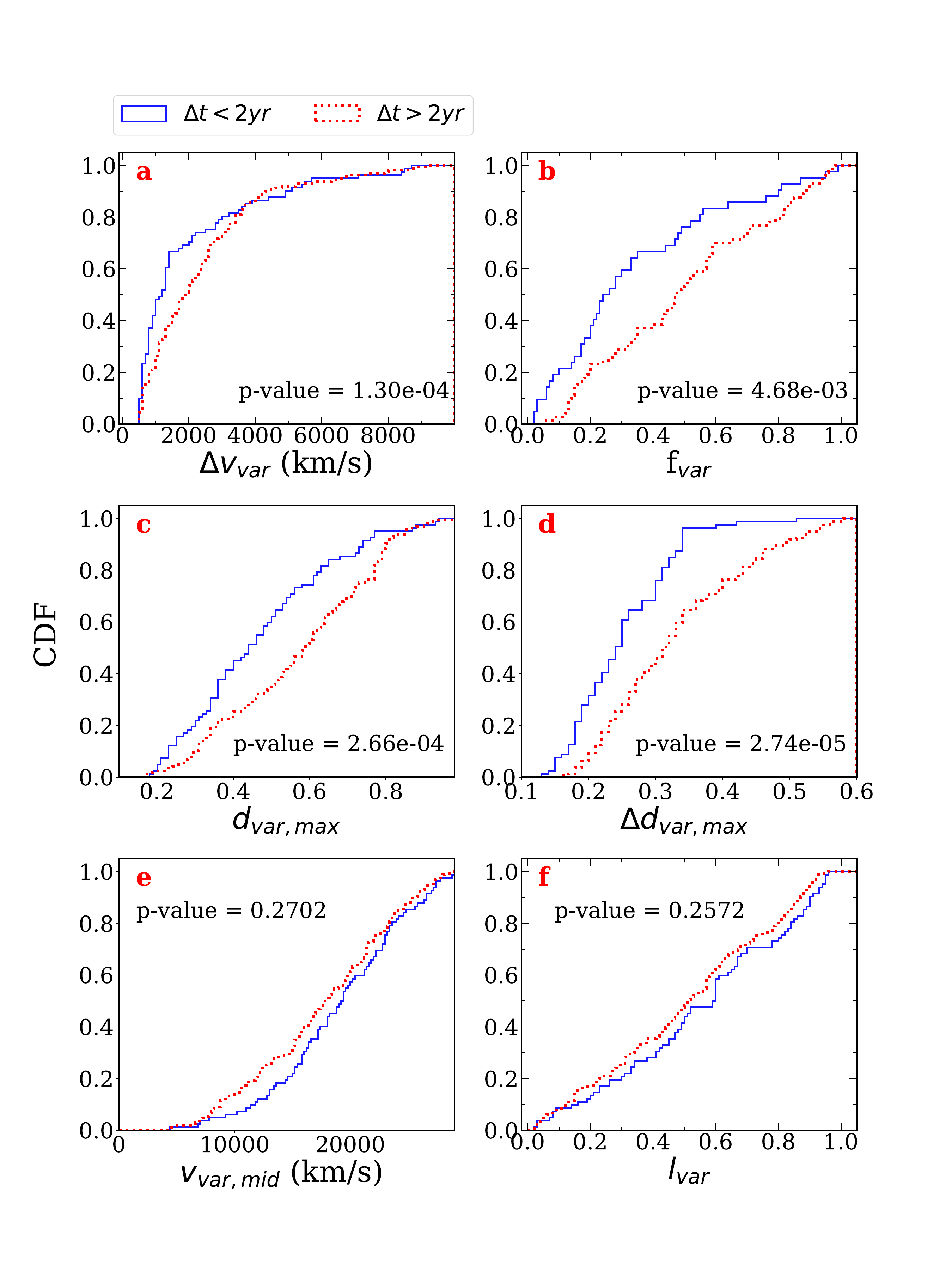}
    \caption{Cumulative distributions of different parameters of the variable regions as discussed in Section~\ref{sec:varregion}. The results are presented for short (blue) and long (red) time-scales using method I (i.e using epoch pairs that show maximum fractional equivalent width variation). In each panel we also provide p-values for the KS-test between the two cumulative distributions. }
    \label{fig:cdf_tdep_max_dw_epochs_method1}
\end{figure}

We compare the cumulative distribution functions (CDF) of all the parameters of variable regions mentioned above in Fig.~\ref{fig:cdf_tdep_max_dw_epochs_method1} for the long and short time-scales using method I.
We also discuss the results from the same analysis using method II in the following sub-sections.

\subsubsection{Velocity width of the variable regions}
\label{subsec:fvar}
%

%
From panel (a) of Fig.~\ref{fig:cdf_tdep_max_dw_epochs_method1} it is clear that the distribution of  $\Delta v_{var}$ is significantly different (p-values for the KS-test are also provided in the figure) between the  short and long time-scales. 
The median $\Delta v_{var}$ is 1236 \kms\ and 2066 \kms\ for short and long time-scales respectively 
The 16th and 84th percentiles in $\Delta v_{var}$ are 618-3775 \kms\ and 823$-$4097 \kms\ respectively for the two time ranges.
These are significantly smaller than the typical velocity widths of the BAL in our sample which has a median of $\sim$ 14000 \kms\ \citep[see Table 1 of ][for velocity width of individual UFO BAL components in our sample]{aromal2023}. 
{\it All this shows that  $\Delta v_{var}$ tends to be larger when the time interval probed is larger. }
This result is confirmed even when we use the distribution of variable regions obtained over both time ranges using method II. 
This conclusion is also strengthened by the width difference distribution plotted in the bottom panel of Fig.~\ref{fig:var_reg_overlap} for the common variable regions.

In comparison \citet{Gibson2008} have found the median width of their variable regions to be $\sim$ 2000 \kms\ over a time-scale of 3-6 yrs. This is consistent with our values for long time scales. In the case of \citet{Filiz2013} the median width of the variable regions is 713.6 \kms which is lower than what we measure. 
The rest frame time-scale probed in their sample varies from 1.0-3.7 yrs. The main reason for the difference could have  come from the definition of the variable region used. 

\subsubsection{Fractional width of the variable regions}
In panel (b) of Fig.~\ref{fig:cdf_tdep_max_dw_epochs_method1}, we plot the CDF of the fraction of variable region ($f_{var}$) for the short and long time scale sub-samples. The median values are 0.25 and 0.48 respectively for  the short and long time scales respectively (see Table~\ref{tab_var_prop}). We see that 
only $\sim$20\% of the BAL troughs have more than 80\% of the velocity range covered by variable regions when we consider the long time scale. This is $\sim$10\% for the short time scale. The KS-test also confirms that the distribution of fractional width of the variable regions for the short and long time scales are significantly different with a p-value of $4.7\times 10^{-3}$. 
We obtain the same result when we repeat this exercise using method II as well.
Thus the above result is not influenced by the time-sampling of the BAL variability.

This is in line with the expectation based on the behavior of $\Delta v_{var}$ 
for ``common" variable regions.
This finding that only a small fraction of BAL regions show significant variability is in agreement with many other studies in the literature \citep{Gibson2008, capellupo2011, Filiz2013}.  However median values of fractional width of variable regions are not readily available in these references for direct comparison with our results. {\it In short, we find that the variable regions mostly occupy only a few tenths of the entire width of the BAL and these $f_{var}$ values in general increase with time.}

\subsubsection{Dependence on the absorption depth}

In \citet{aromal2023} we found that the 
variation of the equivalent width (i.e \fdw) is increasing with the increasing time-scale 
probed. Here, we ask if there is such a trend observed in the case of the variable regions as well.
In panel (c) of Fig.~\ref{fig:cdf_tdep_max_dw_epochs_method1}, we plot the CDF of the maximum depth of the variable regions for two different time ranges considered here. The $d_{var}$ values are found to be smaller (median value of 0.44 and 16th and 84th percentile range of 0.27 to 0.65) in the case of shorter time-scale compared to that (a median value of 0.59 and 16th and 84th percentile range of 0.34 to 0.70) for the longer time-scale. 
The CDF of $d_{var}$ for the two sub-samples are also found to be statistically different (with a p-value of 2.66$\times 10^{-4}$ for the KS-test).  Once again we note that the trend seen here is reproduced even when we use method II. 
This suggests that on average the variable regions identified over a shorter time scale tend to have shallower absorption compared to variable regions identified over a longer time scale. 

In panel (d) of  Fig.~\ref{fig:cdf_tdep_max_dw_epochs_method1} we show the  CDF of maximum change in depth in the variable regions for the two time bins using method I. We notice that the median value of $\Delta d_{var}$ is 0.25 (with a range between 16 and 84 percentiles of 0.18 to 0.33) and 0.32 (with a range between 16 and 84 percentiles of 0.22 to 0.46). It is also clear from the figure that the two distributions are statistically different (for both methods used). 
{\it This confirms that the absorption depth difference in the variable regions is larger when they are detected over long time scales. }
This again is consistent with our findings for the common variable regions (see Fig.~\ref{fig:var_reg_overlap}).

%
%

\subsubsection{On the location of the variable regions}
Next, we look at the CDF of $v_{var,mean}$ and $l_{var}$ to check if the presence of variable regions depends on the absolute velocity with respect to \zem\ and/or its relative position within the trough itself (see panels e and f of Fig.~\ref{fig:cdf_tdep_max_dw_epochs_method1}).
Recall in our sample, the absorption profiles typically sample the velocity range of 4000-28000 \kms with more than 40 quasars contributing to the velocity range 12000-26000 \kms. If the variable regions occur randomly in the velocity space then we expect the median to be around $\sim$19000 \kms.
When we consider the measurements using method I, we notice that the median values of these quantities are slightly higher for short time scales. 
For example, the medians of the central velocity of the variable regions are 19339 \kms\ and 17947 \kms\ for the short and long time scales respectively.
However, the difference between the two distributions is not statistically significant based on the p-values obtained. This is the case even when we use method II.  However, with this method, the median values of the two distributions are nearly similar (i.e. 19493 \kms\ and 19134 \kms\ for short and long time scales respectively). 
These values are close to our expectation above which suggests  that the occurrence of variable regions in velocity space with respect to the BAL QSO is close to a random distribution. 


To explore this further, we study the distribution of $l_{var}$ in panel (f) of Fig.~\ref{fig:cdf_wdep_max_dw_epochs}.
The median values of $l_{var}$ are 0.59 and 0.51 respectively for short and long time scales when we use method I. This indicates a nearly symmetric distribution of variable regions in the velocity space for the long time-scale studied. 
The same for the short time scales may be slightly favoring larger velocities. However, this is not statistically significant as suggested by the  p-values of the KS test.
Indeed, when we consider the measurements based on method II, the median values are identical and the two distributions nearly follow each other with high p-values. The median of $l_{var}$ are 0.55 and 0.55 for short and time-scales (using method II) and they confirm that the locations of the variable region is close to random  (i.e. expected median of 0.5 in that case) within the BAL profile.
{\it Hence, the variable regions show no preference in velocity space which indicates that such large variability may be induced by random processes that can occur anywhere in the flow.}



\subsubsection{Dependence on the sign of variation}
\label{subsec:var_region_signW}

\begin{figure}
    \centering
    \includegraphics[viewport=10 90 2450 1600, width=\textwidth,clip=true]{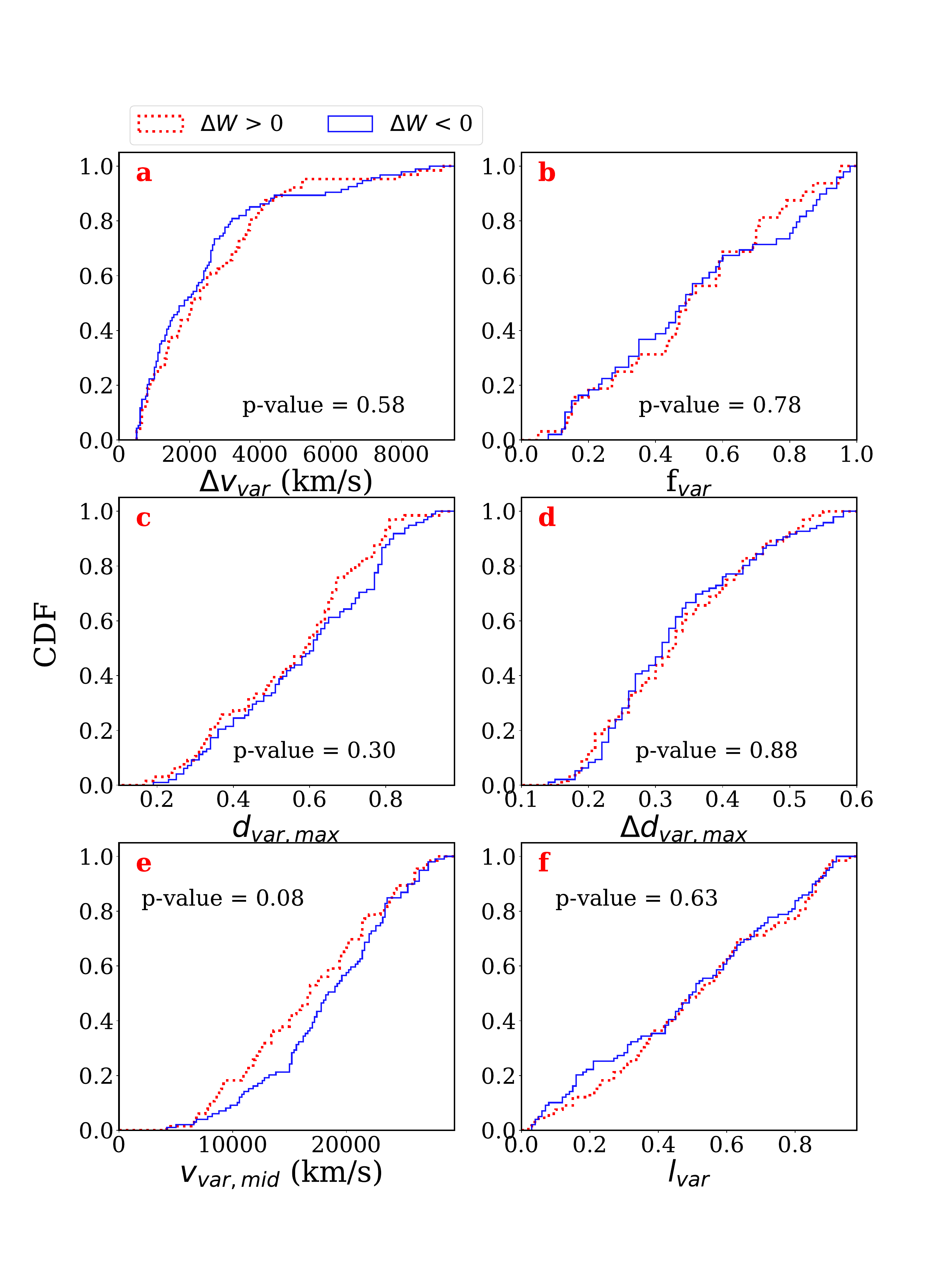}
    \caption{This figure shows the cumulative distribution function of different properties of variable regions for \dw\ > 0 and <0 using epoch pair per source with maximum \afdw\ observed (method I).}
    \label{fig:cdf_wdep_max_dw_epochs}
\end{figure}

In this section, we ask if the properties of the variable regions are different depending on whether the overall absorption is increasing or decreasing in strength. To do that,
we separate the variable regions into two 
sub-samples with one of them coming from epoch pairs that show an increase in absorption and the other showing a decrease in absorption with time. 
%
We then compare the distributions of parameters of the variable regions from these two sub-samples.

Results using method I are summarized in Fig.~\ref{fig:cdf_wdep_max_dw_epochs} where panels are as in Fig.~\ref{fig:cdf_tdep_max_dw_epochs_method1}.
The blue and red curves are for variable regions identified for weakening and strengthening absorption respectively.
%
%
Only $v_{var,mid}$ shows a tentative difference in the distributions as indicated by a p-value = 0.08 when we use method I. However this result is not substantiated by method II for which the p-value is 0.20. 
{\it Hence, we conclude irrespective of whether the overall absorption is increasing or decreasing, the key properties of variable regions remain the same.}



\subsection{Connection between variable regions and \afdw}
  \begin{figure*}
\begin{subfigure}{0.47\textwidth}
    \centering
    \includegraphics[viewport=40 50 1170 1200, width=\linewidth,clip=true]{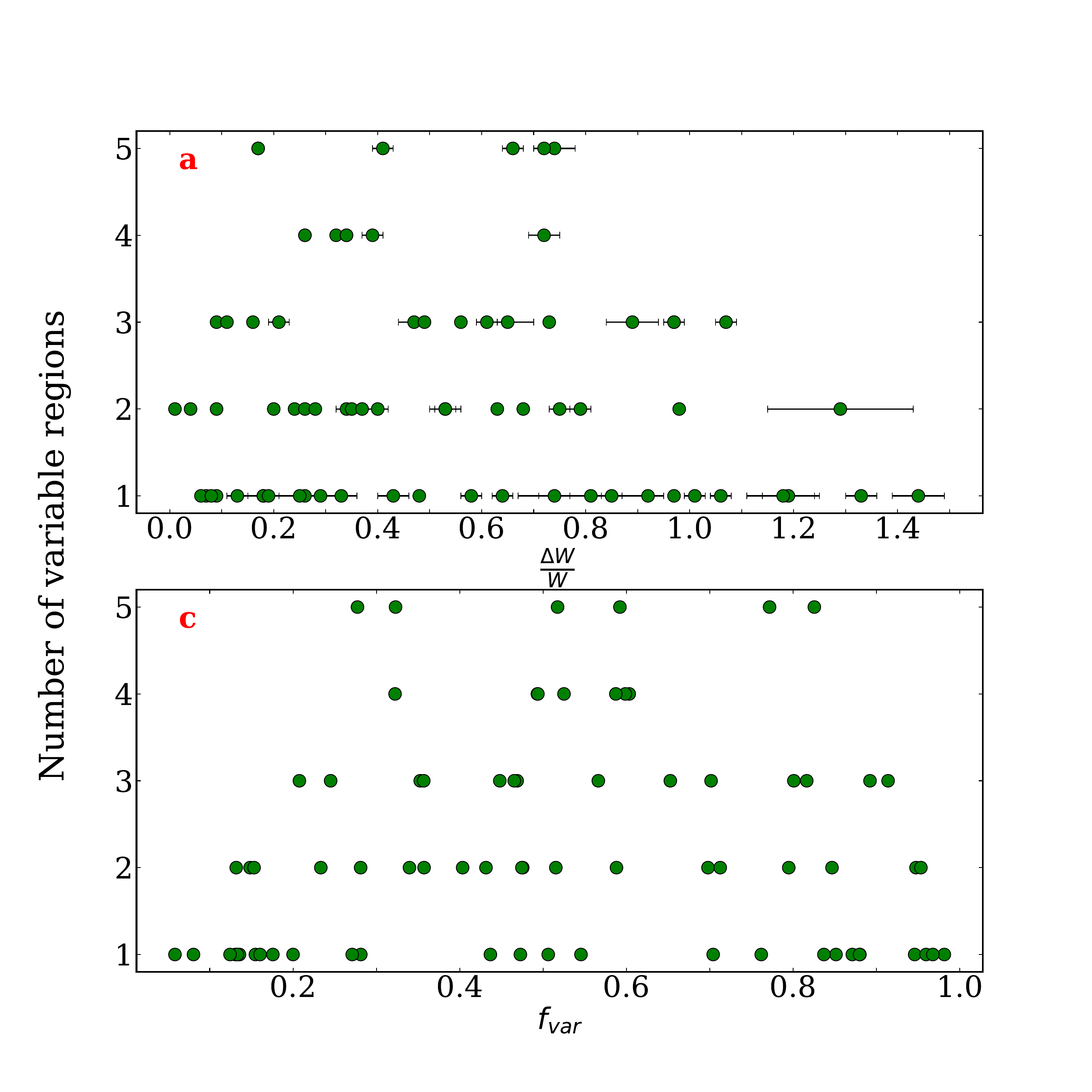}
    \end{subfigure}
    \begin{subfigure}{0.47\textwidth}
    \centering
      \includegraphics[viewport=40 50 1170 1200, width=\linewidth,clip=true]{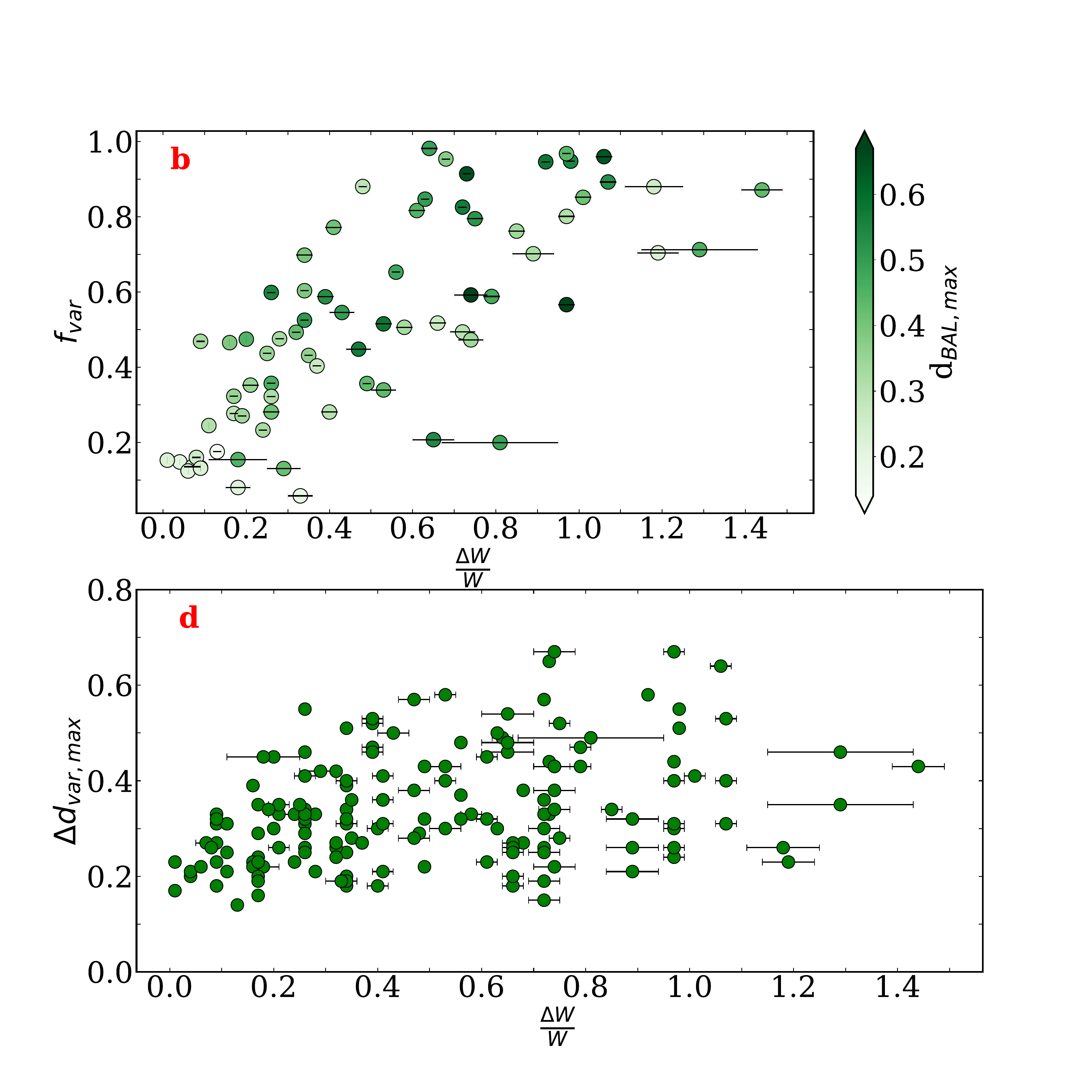}
      \end{subfigure}
    \caption{
    Panel (a) shows the number of variable regions (as derived from method I) as a function of the maximum fractional change in the equivalent width of the BAL ($\frac{\Delta W}{W}$).
    Panel (b) shows $f_{var}$ as a function of $\frac{\Delta W}{W}$ and the points are color-coded with the maximum depth variation in the BAL denoted as d$_{BAL,max}$.
    In panel (c),  the number of variable regions is plotted with respect to $f_{var}$.
    Panel (d) shows $\Delta d_{var,max}$ as a function of $\frac{\Delta W}{W}$.
    }
    \label{fig:wvsvr}
\end{figure*}

\begin{figure}
    \centering
    \includegraphics[viewport=50 0 2400 770, width=\textwidth,clip=true]{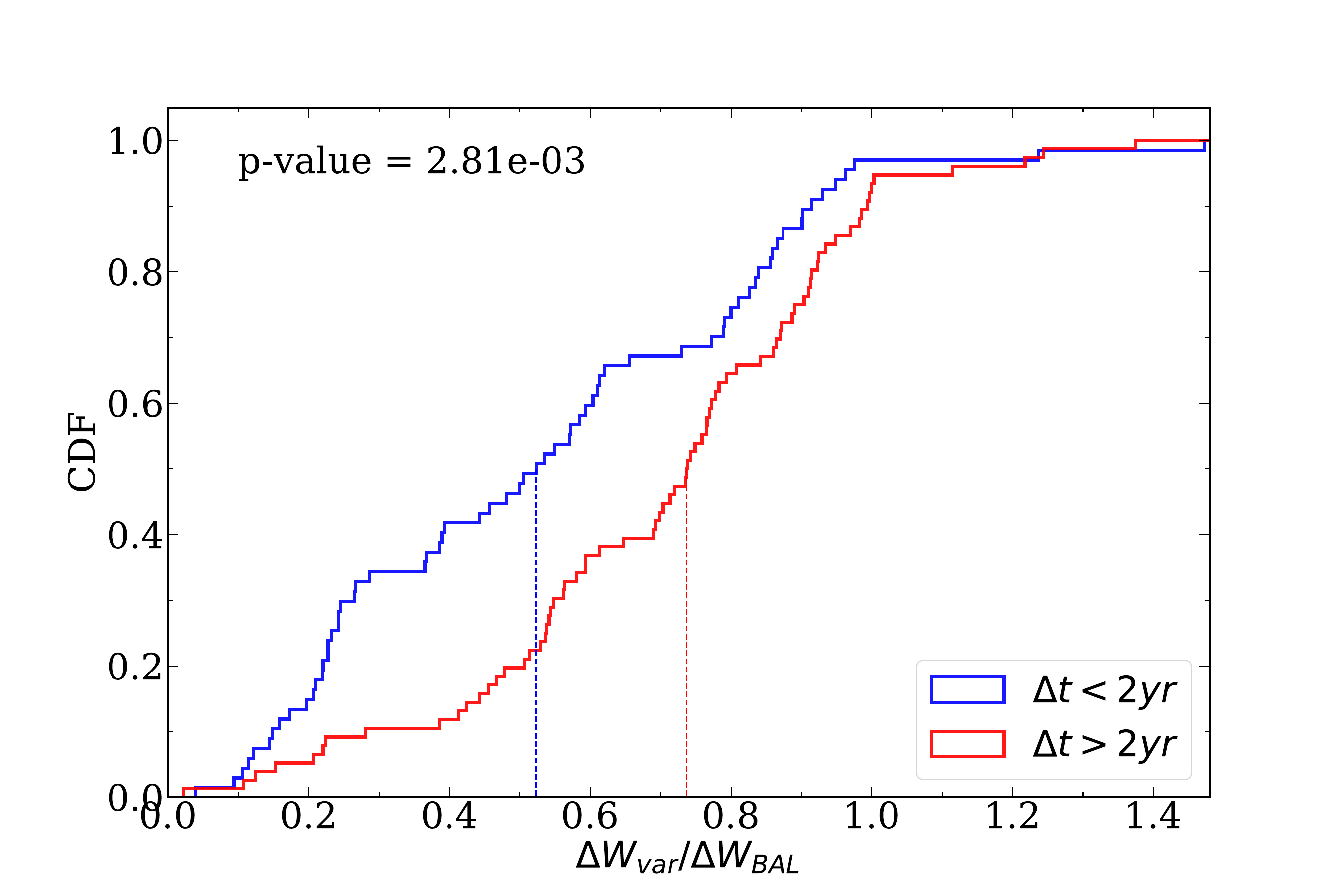}
    \caption{Cumulative distributions of the ratio of  $\Delta W_{var}$ contributed by all the variable region inside the BAL and 
    $\Delta W_{BAL}$ for the entire BAL for $< 2$ yr (blue) and $> 2$ yr (red) time scales.
    The median values of this quantity are shown in dotted lines in their respective colors.
    The p-value from the KS test is also shown in the figure indicating the distributions are statistically different. }
    \label{fig:dw_var_by_fdw_bal}
\end{figure}

\begin{figure}
    \centering
    \includegraphics[viewport=50 0 2400 770, width=\textwidth,clip=true]{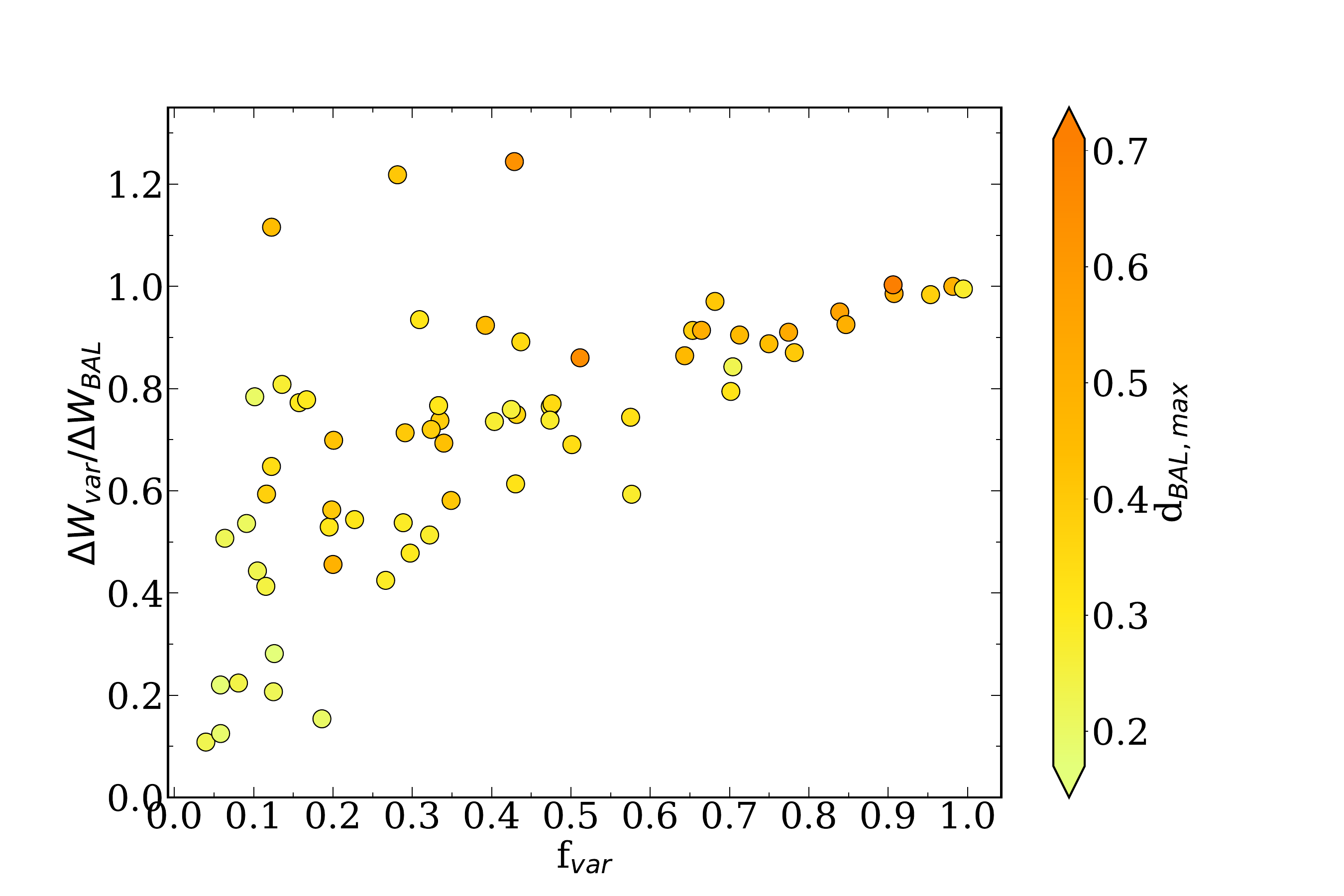}
    \caption{This figure shows $\Delta W_{var}$/$\Delta W_{BAL}$, ratio of the equivalent width variations
    contributed by the variable regions and the entire BAL, as a functin of $f_{var}$
   for $\Delta t >$ 2 yr. 
   The points are color-coded with the maximum depth variation in the BAL denoted as d$_{BAL,max}$.}
    \label{fig:fvar_vs_fdw_var}
\end{figure}

In this section, we explore how the variable regions that show significant absorption line variations over a narrow velocity range influence the global absorption line variability quantified by the \afdw. 
In panel (a) of Fig.~\ref{fig:wvsvr} we plot \afdw\ vs the number of variable regions identified between two epochs that show maximum \afdw\ for each BAL component (i.e method I). The only trend that is evident is that the large \afdw\ are not driven by the presence of a large number of variable regions. 
Recall \afdw$\ge0.67$ will correspond to a factor 2 variations in the equivalent width \citep[called as "highly variable" BAL in][]{aromal2023}.  It is clear from the plot that such large variations in \civ\ equivalent widths are driven most of the time by less than 3 variable regions.  It is also evident from the figure that BALs that show $\ge$ 4 variable regions tend to show \afdw\ less than 0.8. {\it Thus it appears that the number of variable regions does not affect the measured equivalent width variations.}


In panel (b) of Fig.~\ref{fig:wvsvr} we plot \afdw\ vs $f_{var}$. We find a clear correlation between the two quantities. The Spearman rank correlation coefficient is 0.7762 and the p-value is $<<10^{-5}$. It is evident that the ``highly variable" BALs tend to have $f_{var}>0.5$.  {\it Combining this with the lack of correlation between \afdw\ and the number of variable regions, we can conclude that the large equivalent width variations are primarily driven by a small number of variable regions having larger velocity widths.}

Note $f_{var}$ is computed taking into account the total velocity range covered by all the variable regions present. Therefore, it will be interesting to see if there is any correlation between $f_{var}$ and the number of variable regions.
%
This is what we plot 
in panel (c) of Fig.~\ref{fig:wvsvr}.
We do not find any clear trend between the two quantities.  
If we consider $f_{var}>0.6$, that predominantly contains ``highly variable" BALs (see panel b of Fig.~\ref{fig:wvsvr}), 61.5\% of the cases have less than two variable regions, 23\% of the cases have three variable regions and the remaining 15.4\% region have four or five variable regions.

%
We can conclude that the BAL troughs with large equivalent width variations ('`highly variable" BALs)
have a small number of variable regions with larger velocity widths (i.e. typically covering 50 to 100\% of the BAL profile) which means that the whole BAL profile is varying. 
On the contrary BALs with \afdw$<$0.3 tend to have $f_{var}<0.5$ and in that case, observed equivalent width variations are driven by narrow variable regions.

%

In panel (d) 
of Fig.~\ref{fig:wvsvr} we plot the maximum depth variation vs. \afdw.  The Spearman's rank correlation coefficient is, $r_s = 0.41$,  with a p-value of $5.22\times 10^{-8}$. However, it is evident in this plot that when \afdw\ is more than 0.2, the  correlation between the two quantities becomes significantly weaker (i.e. $r_s=0.2$ and p-value of 0.01). Thus the correlation found for the full sample mainly reflects the fact that for \afdw\ $<$ 0.2, we do not have $\Delta d_{var,max}$ more than 0.35. 
If we focus on the ``highly variable" BAL components, for a given \afdw, $\Delta d_{var,max}$ shows a large scatter. This once again confirms the importance of $f_{var}$ in governing the large \civ\ equivalent width variations.

\subsubsection{Relative contribution of variable regions to global BAL variations}
Next, we quantify the relative contribution of variable regions to the total BAL variations in terms of equivalent width changes, \dw.
For each consecutive epoch pair with variable regions detected, we estimate the sum of the absolute value of \dw\ contributed by all the variable regions ($\Delta W_{var}$) and divide it by $\Delta W_{BAL}$ measured
over the entire BAL region. 
As we are using absolute values the ratio can be more than 1 if there are uncorrelated variable regions (i.e. some part of the absorption shows an increasing trend while the other part shows a decreasing trend). Examples of such variability are discussed in Section~\ref{subsec:uncorrelated_var}.
The cumulative distribution function of this quantity (i.e. $\Delta W_{var}/\Delta W_{BAL}$) for short ($< 2$ yr) and long ($> 2$ yr) time scales are shown in Fig~\ref{fig:dw_var_by_fdw_bal}. The distributions for the different time scales are statistically different as suggested by the p-value = 0.002 from the KS test.  
The median of ${\Delta W_{var}}/{\Delta W_{BAL}}$ for short and long time scales are 0.53 and 0.74 respectively.
This means that 
in 50\% of the BAL troughs the equivalent width changes in the variable regions account for more than 74\% (respectively 53\%) of the total observed \civ\ equivalent width variability for long (respectively short)
time-scales.
We see that for long time scales, more than 50$\%$ of the equivalent width variations are contributed by variable regions in 80$\%$ of the BAL troughs whereas the same is true for 54$\%$ of the cases in short time scales.
This indicates that the variable regions contribute a large fraction of the BAL variations, especially at long time scales compared to the short time scales.

We also looked at how $\frac{\Delta W_{var}}{\Delta W_{BAL}}$ depends on $f_{var}$ at long time scales ($\Delta t >$ 2 yr) as shown in Fig~\ref{fig:fvar_vs_fdw_var}.
At extreme ends, i.e. at $f_{var} < 0.1$ and $> 0.6$, $\frac{\Delta W_{var}}{\Delta W_{BAL}}$ has relatively smaller and higher values respectively as expected.
Even though this leads to high Spearman's correlation coefficient (r = 0.67), we notice that at intermediate values of f$_{var}$, there is a significant scatter in the relation 
(r = 0.27 for $0.1 <$ f$_{var} < 0.6$).
This means at these values, $\frac{\Delta W_{var}}{\Delta W_{BAL}}$ is independent of f$_{var}$.
We also note that for this range in f$_{var}$, there is no clear trend between $\frac{\Delta W_{var}}{\Delta W_{BAL}}$ and velocity width ($\Delta v_{var}$) or maximum depth change ($\Delta d_{var,max}$) of the variable regions.  
{\it Combining this with the fact that a large fraction of the BAL variations are contributed by the variable regions indicates that the significant BAL variations are mostly localized to variable regions irrespective of their extent in velocity space.
From Section~\ref{subsec:fvar}, we know that the median f$_{var}$ values 
are in the range 0.25-0.48
which implies most of the BAL variations do happen in comparatively smaller regions.
This again proves the importance of small-scale inhomogeneities in the physical parameters of the flow in driving the local optical depth variations inside the BAL.}

\subsection{Is there any uncorrelated variability in the UFO sample ?}
\label{subsec:uncorrelated_var}
\begin{figure*}
    \centering
    \includegraphics[viewport=72 60 2200 1170, width=2\textwidth,clip=true]{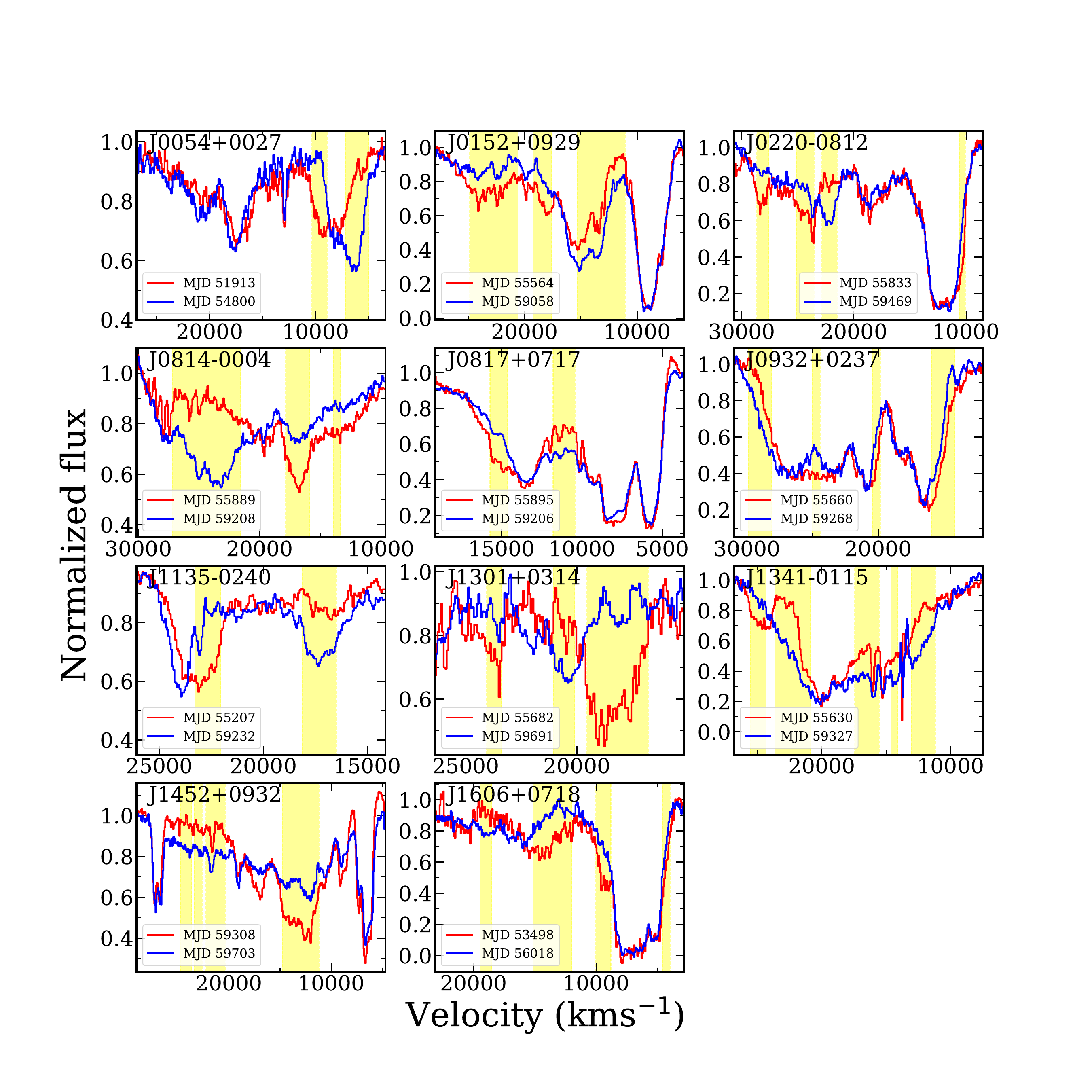}
    \caption{
    This figure shows the uncorrelated BAL variations in 11 UFO BAL sources in our sample where the two spectra of the pairs of epochs are shown in velocity space with blue and red colors. Each subplot shows the BAL regions of the respective source and the variable regions with uncorrelated variations for the epoch pairs are shown in yellow shaded area.
    }
    \label{fig:uncorrelated_var_all}
\end{figure*}

The absorption line variability in BAL is known to correlate over large velocity scales \citep[see for example objects discussed in][]{Aromal2021,aromal2022}. Such correlated absorption line variability is usually attributed to photo-ionization-induced variability. We also notice that in our sample, when multiple BAL components are present in a single quasar, we do see a correlation in the equivalent width variability between these components \citep[see section 4.6 in][]{aromal2023}. 
However, if the variable regions are governed by local changes in physical conditions, then we expect their absorption variability to not correlate with that of the rest of the profile.

Here, we search for any uncorrelated variability (i.e. different variable regions showing uncorrelated variability) within an identified BAL component by considering all the consecutive epoch pairs of each quasar and looking for at least one pair of variable regions with uncorrelated variability. Out of 64 UFO BAL quasars in our sample, we find 11 of them (i.e $
\sim 17$\%) show uncorrelated variability within their profile. The epoch pairs with such variations are shown in Fig.~\ref{fig:uncorrelated_var_all}. 

\begin{figure}
    \centering
    \includegraphics[viewport=50 20 2420 800, width=\textwidth,clip=true]{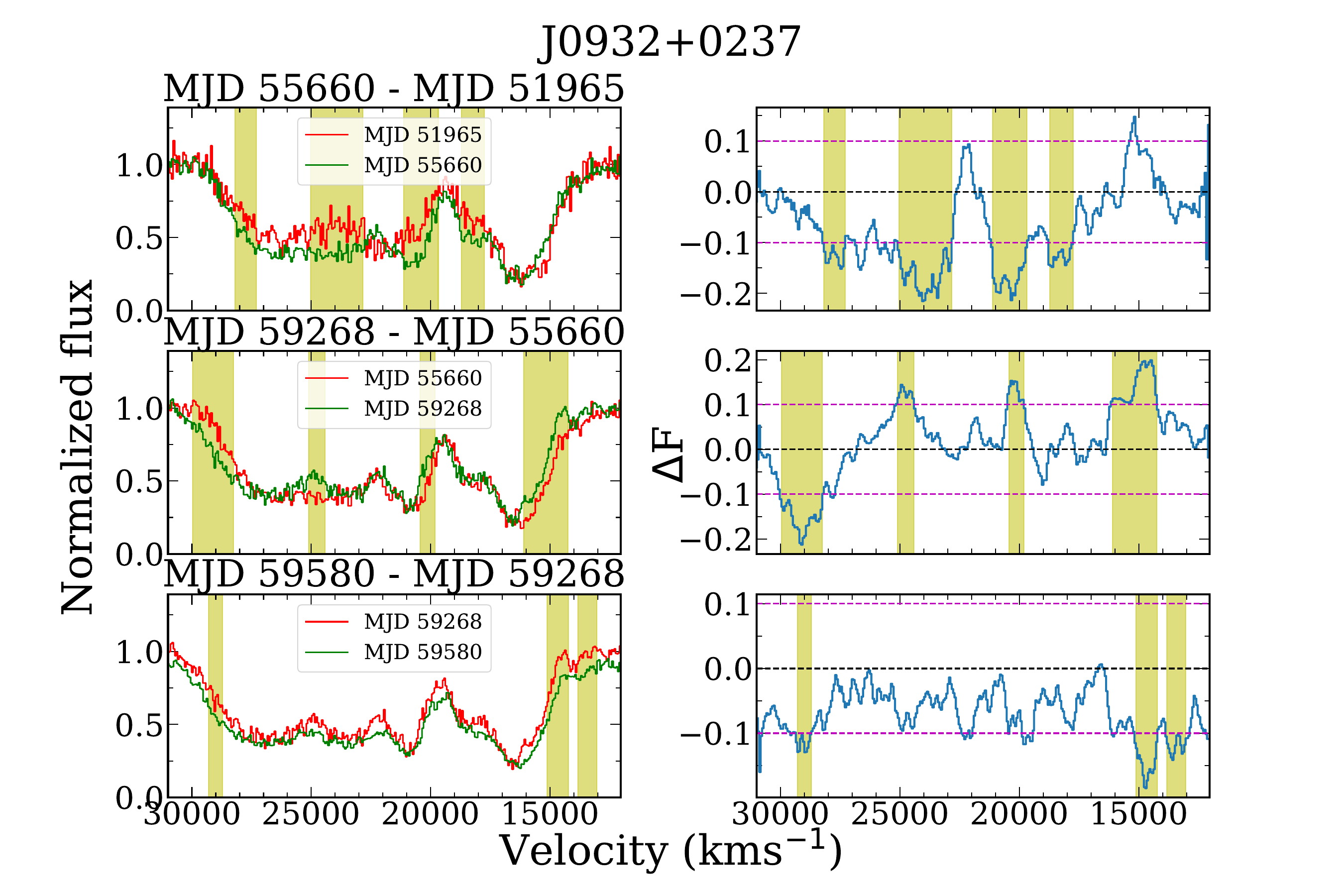}
    \caption{Same as Fig~\ref{fig:eg_var_region} but for J0932+0237.
    The variable regions are shown as yellow-shaded regions
    inside the BAL region.
    Spectra are shown in the left panels as red and green curves
    respectively for two consecutive pairs of epochs. The corresponding flux differences ($\Delta$F)
    are given in the right panels.
    }
    \label{fig:eg_uncorr}
\end{figure}

When we carefully analyze all the available spectra we notice that the uncorrelated variability 
occurs due to (1) the appearance and/or disappearance of a new velocity component during any particular epoch and (2) an apparent shift in velocity for an already existing absorption component. 
In nine cases the uncorrelated variability measured is mainly due to the appearance or disappearance of a new component. In six cases 
(J0054+0027, J0220-0812, J0932+0237, J1135-0240, J1452+0932 and J1606+0718) we do see a shift in the velocity of some components. Note in all cases the profiles are also considerably changed 
so that the strict definitions of kinematic shift as explained in \citet{grier2016} may not be applicable in these cases.
%

Uncorrelated variability can occur 
when the density fluctuates within the region producing the BAL absorption.
In this case, the ionization state of the gas may change from one place
to the other.
It would be thus quite interesting to perform observations
over a larger wavelength range to follow the variations 
of different species to study the ionization changes within the flow.
We also note that almost all of these uncorrelated variations occur at time scales of more than two years in the rest frame which may also correspond to the time scales of the lifetime of these density fluctuations inside the flow.

Interestingly when multiple epoch observations are available, in 10 out of 11 cases we see correlated variability before and after the epoch pairs that show uncorrelated variability (See for example Fig~\ref{fig:eg_uncorr}). Only in one case (i.e. J1452+0932) we see uncorrelated variability in all epoch pairs considered. In this case, we have 4 epochs and the total \civ\ equivalent width shows a monotonous decrease (by a factor $\sim$2 compared to that measured in the first spectrum) with time.

{\it In summary, the uncorrelated variability in $\sim$17\% of our UFO sample is mainly dominated by a readjustment in the gas distribution along our line of sight. }
Future follow-up observations will be important to understand these uncorrelated variations better.

\section{Conclusions}
\label{sec:conclusions}

In this work, we have presented the absorption depth based analysis of time variability of the \civ\ absorption from 80 distinct UFO BAL components observed in the spectra of 64 UFO BAL quasars. 

To start with, we split the BAL regions in each source into velocity bins of 2000 \kms\ width and calculate the equivalent width for each bin in order to study the dependence of variability on outflow velocity ($v_{out}$).
We find that the strength of variability as quantified by $|\frac{\Delta W^v} {W^v}|$ increases significantly with increasing outflow velocity. This result is found to be independent of the choice of bin width and consistent with what has been found in \citet{aromal2023} using equivalent widths of BAL troughs.

For each source in our sample, we also identify ``variable regions" as those regions inside the BAL trough where the flux difference ($\Delta F = F_2-F_1$ ) between two epochs is greater than 0.1 over a velocity width of at least 500 \kms\ using two methods
as explained in Section~\ref{sec:varregion}. 
In the present sample, 61 of the 64 UFO BAL quasars show 
BAL equivalent width variations at $>3\sigma$ significance level.
For these BAL QSOs showing significant equivalent width variations, we note that 60 of the 61 variable BALs do show at least one detectable variable region.
%
We see that the number of variable regions in two time intervals [i.e $<$2 yrs (referred to as short) and $>$2 yrs (referred to as long)] are 82 (from 42 BALs) and 166 (from 73 BALs) respectively.
This indicates that
even though the number of variable regions per BAL is $\sim 2$ irrespective of the time scale, the total number of BALs having variable regions increases with time as it almost doubled for the time scales we have considered.

We also looked at 38 BALs for which we identified ``common" variable regions that are present in  both the short and long time scales.
We found that even though these variable regions do not show considerable change in their central velocities, the velocity width increases significantly with longer time scales.

To characterize the variable regions in terms of their width, depth, and position for the full sample, we define six parameters, namely $\Delta v_{var}$, f$_{var}$, $d_{var,max}$, $\Delta d_{var,max}$, $v_{var,mid}$ and $l_{var}$ as explained in Section~\ref{sec:varregion}.
We find that the widths of the variable regions ($\Delta v_{var}$) are generally only a few thousand \kms\ which is much smaller than the typical velocity width of the BAL and this results in the ratio between these two widths  (f$_{var}$) to be a few tenths in most cases. We note that the typical velocity width of variable regions is similar to that of the mini-BALs \citep{hamann2004}.
This indicates that most of the significant variations occur over narrower regions as opposed to the entire BAL region.
Both $\Delta v_{var}$ and f$_{var}$ are highly dependent on the time scales 
and increases with time 
between short and long time scales.
Similarly, the quantities related to the absorption depth in the variable regions, \ie\ both the maximum depth ($d_{var,max}$) and the maximum change in the depth ($\Delta d_{var,max}$) of the variable regions, show larger values when they are detected over long time scales.
The distributions of the mid-velocity ($v_{var,mid}$) and the relative position within the BAL ($l_{var}$) of the individual variable regions indicate that the occurrence of variable regions in velocity space does not show any preference with respect to \zem\ and also on its relative position within the trough itself and hence turns out to be closer to a random distribution.
As a consequence, they also show similar distributions irrespective of the variability time scales.
This suggests a possible presence of local instabilities throughout the flow.

We also investigated if the properties of the variable regions are different depending on whether the overall absorption is increasing or decreasing in strength.
We find key properties of variable regions remain the same irrespective of this difference.

Next, we looked at how the relatively narrow variable regions affect global absorption line variability parameters such as the fractional BAL equivalent width variations (\afdw).
The lack of correlation between the number of variable regions and \afdw\ combined with the observed strong correlation between $f_{var}$ and \afdw\ indicate that the large equivalent width variations are primarily driven by a small number of variable regions having larger velocity widths. This underlines the importance of $f_{var}$ in determining the strength of total BAL variability.

We find that, at long  (respectively short) time scales,
in 50\% of the BAL troughs, the equivalent width changes in the variable regions account for more than 75\% (respectively 53\%) of the total observed \civ\ equivalent width variability of the BAL trough.
We see that for long time scales, more than 50$\%$ of the equivalent width variations are contributed by variable regions in 80$\%$ of the BAL troughs whereas the same is true for 54$\%$ of the cases in short time scales. 

Finally, in 17$\%$ of the sources (11 out of 64), 
we see different parts of the same BAL vary in opposite directions (what we call uncorrelated variability).
A detailed analysis of the nature of these variations indicates that they may be dominated by a readjustment in the gas distribution along our line of sight.
It would be thus quite interesting to perform observations
over a larger wavelength range to follow the variations 
of different species to study the ionization changes within the flow.

Simple radiatively driven disk wind models \citep{murray1995}, in general, assume a smooth radial density field along the flow.
However, hydrodynamical simulations \citep{proga2000, Dyda2018} indicate that there can be considerable density enhancements in the flow that will lead to non-smooth density profiles whose time evolution can lead to variability signatures.
It is tempting to say that the variable regions detected in our sample may originate from such deviations in an otherwise smooth density profile and contribute significantly to the overall BAL variability.
Combining this with our finding that there is no velocity preference for the occurrence of variable regions inside BAL, we require these density variations to be randomly distributed along the flow.

Now, such density fluctuations can be due to : (1) Kelvin-Helmholtz instability between different parts of the outflow which occurs over a few years time scales as noted in \citet{proga2000}, (2) injection/disappearance of new gas components in our LOS, (3) evaporation of gas \citep{dyda2020} and (4) formation of clumps due to cooling instabilities. The large velocity width of variable regions found here will provide interesting constraints on these possibilities. Further, we notice an increase in the width of the variable regions over longer time scales in a majority of the sources suggesting the fluctuations may grow over time affecting much of the BAL profile.
We note that a lot more insights can be gained using follow-up high-resolution spectroscopy of our sources (that will resolve the variable regions better) with better time sampling. In particular, probing the ion ratio variations in the variable regions will be very important to get more handle on the origin and evolution of the variable absorption regions in the BAL flows.

\section*{Acknowledgements}
PA thanks Labanya K Guha for helpful discussions on several Python programming techniques used in this paper.
PPJ thanks Camille No\^us (Laboratoire Cogitamus) for 
inappreciable and often unnoticed discussions, advice and support. PPJ is partly supported by the Agence Nationale de la Recherche under contract HZ-3D-MAP, ANR-22-CE31-0009.

\section*{Data Availability}
Data used in this work are obtained using SALT. Raw data will become available for public use 1.5 years after the observing date at https://ssda.saao.ac.za/.



\bibliographystyle{mnras}
\bibliography{mybib_bal} 


\appendix

\bsp	
\label{lastpage}
\end{document}